# A Deficiency Problem of the Least Squares Finite Element Method for Solving Radiative Transfer in Strongly Inhomogeneous Media

J.M. Zhao[a], J.Y. Tan[b], L.H. Liu[a]*

[a] *School of Energy Science and Engineering, Harbin Institute of Technology, 92 West Dazhi Street, Harbin 150001, People's Republic of China*

[b] *School of Auto Engineering, Harbin Institute of Technology at Weihai, 2 West Wenhua Road, Weihai 264209, People's Republic of China*

**Abstract**

The accuracy and stability of the least squares finite element method (LSFEM) and the Galerkin finite element method (GFEM) for solving radiative transfer in homogeneous and inhomogeneous media are studied theoretically via a frequency domain technique. The theoretical result confirms the traditional understanding of the superior stability of the LSFEM as compared to the GFEM. However, it is demonstrated numerically and proved theoretically that the LSFEM will suffer a deficiency problem for solving radiative transfer in media with strong inhomogeneity. This deficiency problem of the LSFEM will cause a severe accuracy degradation, which compromises too much of the performance of the LSFEM and makes it not a good choice to solve radiative transfer in strongly inhomogeneous media. It is also theoretically proved that the LSFEM is equivalent to a second order form of radiative transfer equation discretized by the central difference scheme.

---

* Corresponding author. Tel.: +86-451-86402237; fax: +86-451-86221048.

*Email addresses:* jmzhao@hit.edu.cn (J.M. Zhao), tanjy@hit.edu.cn (J.Y. Tan), lhliu@hit.edu.cn (L.H. Liu)

## Nomenclature

| | |
|---|---|
| $\mathbf{B}$ | Tool matrix defined in Eq. (**12**a) |
| $\mathbf{b}$ | Tool vector defined in Eq. (**12**b) |
| $\mathbf{D}$ | Differential matrix defined in Eq. (**12**d)-(h) |
| $E$ | Relative error |
| $G$ | Incident radiation, W/m$^2$ |
| $\mathbf{h}$ | Vector defined in Eq. (**9**) |
| $I$ | Radiative intensity, W/m$^2$sr |
| $\mathbf{K}$ | Stiff matrix defined in Eq. (**9**) |
| $L$ | Side length, slab thickness, operator defined in Eq. (1a) |
| $\mathbf{n}$ | Normal vector |
| $N_{el}$ | Total number of elements |
| $N_{sol}$ | Total number of solution nodes |
| $q$ | Radiative heat flux, W/m$^2$ |
| $R$ | Residual |
| $s$ | Ray trajectory coordinates |
| $\mathbf{s}$ | Vector defined in Eq. (**12**c) |
| $S$ | Source function defined in Eq. (1b) |
| $T$ | Temperature, K |
| $U$ | Unit step function |
| $V$ | Solution domain |
| $\mathbf{x}$ | Spatial coordinates vector |
| $x, y, z$ | Cartesian coordinates |
| $\beta$ | Extinction coefficient, $m^{-1}$ |
| $\delta$ | Kronecker delta |
| $\varepsilon_w$ | Wall emissivity |
| $\phi$ | FEM shape function |
| $\varphi$ | Azimuth angle |
| $\Phi$ | Scattering phase function |
| $\kappa_a$ | Absorption coefficient, $m^{-1}$ |
| $\bar{\kappa}_a$ | A constant parameter indicating the strength of extinction |
| $\kappa_s$ | Scattering coefficient, $m^{-1}$ |
| $\mu, \eta, \xi$ | Cartesian components of $\mathbf{\Omega}$ |
| $\varpi, \bar{\varpi}$ | Frequency domain variable, reduced frequency |
| $\theta$ | Zenith angle |
| $\sigma$ | Stefan-Boltzmann constant, $W/m^2K^4$ |
| $\tau$ | Optical thickness |
| $\tau_\Delta$ | Grid optical thickness |
| $\omega$ | Single scattering albedo |
| $\mathbf{\Omega}, \mathbf{\Omega}'$ | Vector of radiation direction |
| $\Omega, \Omega'$ | Solid angle |

*Subscripts and Superscripts*

| | |
|---|---|
| 0 | Inflow boundary |
| $b$ | Black body |
| $E$ | Exact solution |
| $g$ | Medium |
| $i, j$ | Spatial solution node index |
| $m$ | The *m*th angular direction |
| $w$ | Value at wall |

## 1. Introduction

Radiative transfer is important in many scientific and engineering disciplines, such as combustion[1, 2], remote sensing [3, 4] and optical tomography [5-8], etc. Because the density distribution of media is often non-uniform, the media radiative properties are also inhomogeneous. Furthermore, in optical tomography, the different part of tissues may have distinctly different properties. In this case, radiative transfer in medium with strong inhomogeneity has to be considered. The radiative transfer equation (RTE) have to include a complicated extinction coefficient function, such as, a function with discontinuity. This complicates the numerical solution of the radiative transfer equation. The numerical characteristics of the different numerical methods to solve the RTE with discontinuous extinction coefficient have not been well studied.

Up till now, many numerical methods that are capable of solving radiative transfer in inhomogeneous absorbing, emitting and scattering media have been developed, such as the statistical method, e.g., the Monte Carlo method [9-11], and the deterministic methods based on discretization of the RTE, e.g., the discrete ordinate methods (DOM) [12-14], the finite volume method (FVM) [15-17] and the finite element method (FEM) [18-27], etc. The Monte Carlo method is by far the most versatile but it is very time consuming and is not a good choice for computational intensity applications, such as in optical topography and in solving other inverse radiative transfer problems. Currently, the DOM and the FVM are two popular numerical methods that are versatile and efficient to solve engineering radiative transfer problems. The DOM is based on the finite difference method in spatial discretization which is difficult to be applied to problems with complex geometry. The FVM and the FEM are superior to the DOM for that they can be very conveniently applied to complex geometry. Hence the FEM is very promising to be an alternative robust tool in solving radiative transfer problems with complex geometry.

The schemes of FEM already studied for solving radiative transfer problems including the standard Galerkin scheme [18, 19] and some advanced schemes, such as the least squares scheme [20-23], the streamline upwinding scheme [24, 25], the scheme based on the even-parity form of the RTE (EPRTE) [26] and the scheme based on the second order radiative transfer equation (SORTE) [27]. The FEM based on the Galerkin scheme (GFEM) can be considered as the most basic version of FEM, which has been widely used in other disciplines [28-30]. Former studies [23, 27] reported that the GFEM is instable in solving some radiative transfer problems and will produce results with spurious oscillations. An interpretation to this instability is based on an analogy with the solution of the convection diffusion equation. The RTE is in a form as a special kind of convection-dominated equation without the diffusion term and the convection-dominated property will cause the instability of numerical results of the GFEM [31]. Though in fact the RTE is rather different from the convection diffusion equation because of the exist of an extinction term, this analogy brings an initial understanding of the instability problem. The stability problem of the GFEM in solving the RTE have not been well studied in a theoretical framework. The FEM based on the other advanced schemes mentioned before owns good numerical stability and can overcome the stability problem of the GFEM in solving radiative transfer. However, not all the schemes are ideal for analyzing radiative transfer in inhomogeneous media. The streamline upwinding scheme introduces an undetermined tuning parameter that is not easy to be chosen. The FEM based on the EPRTE and the SORTE [27] have singularity problems in dealing with inhomogeneous media where some locations have very small/zero extinction coefficient due to the exist of the reciprocal of extinction coefficient in the formulation. The LSFEM has not all these problems.

The LSFEM was traditionally believed to be stable and accurate for solving partial differential equations in other disciplines [32-34], and hence for solving radiative transfer problems [20-23]. That is because the LSFEM owns two distinct advantages, such as (1) it is based on the principle of functional minimization, and (2) the stiff matrix produced by the LSFEM is symmetric and positive definite, which is a very good numerical property. Furthermore, it shows both accuracy and stability in many numerical tests [20-23]. It is noted that most of the tests are of radiative transfer in homogeneous media.

Though currently the LSFEM has been successfully applied to solving radiative transfer in homogeneous media with limited numerical verifications, the accuracy and stability of the LSFEM have not been well investigated theoretically. Furthermore, the performance of the LSFEM in dealing with media of discontinuous extinction coefficient and strong inhomogeneity has not even been well studied numerically. In fact, the LSFEM suffers a particular deficiency problem in solving radiative transfer in media with strong inhomogeneity as will be demonstrated in this paper, which will cause severe accuracy degradation.

In this paper, we report an observed deficiency problem of the LSFEM in dealing with radiative transfer in media with strong inhomogeneity. To interpret this deficiency problem, the accuracy and the stability of the LSFEM and the GFEM for solving radiative transfer in homogeneous and inhomogeneous media are studied theoretically via a frequency domain technique. The derived theoretical relations are further verified numerically with critical test cases.

The paper is organized as follows. Firstly, the detailed discretization of the LSFEM and the GFEM is presented. Then the accuracy and stability characteristics of the LSFEM and the GFEM are analyzed by a frequency domain technique and the deficiency problem of the LSFEM is reported and analyzed theoretically. Finally, two critical test cases of one- and two-dimensions are used to verify the theoretical relations and give further understanding the deficiency problem of the LSFEM.

## 2. The LSFEM Discretization of the RTE

### 2.1. Discretization

The RTE for radiative transfer in an absorbing, emitting and scattering medium can be written as

$$L[I] = S \tag{1}$$

in which the linear operator $L[\bullet]$ and the source term $S(\mathbf{x},\mathbf{\Omega})$ are defined as

$$L[I] = \mathbf{\Omega} \bullet \nabla I + \beta I \tag{1a}$$

$$S(\mathbf{x},\mathbf{\Omega}) = \kappa_a I_b + \frac{\kappa_s}{4\pi}\int_{4\pi} I(\mathbf{x},\mathbf{\Omega}')\Phi(\mathbf{\Omega}',\mathbf{\Omega})\,\mathrm{d}\Omega' \tag{1b}$$

where $\mathbf{\Omega}$ is the vector of transfer direction, $\beta = \kappa_a + \kappa_s$ is the extinction coefficient, $\kappa_a$ and $\kappa_s$ is the absorption and scattering coefficient, respectively, $I_b$ is the black body radiative intensity and $\Phi$ is the scattering phase function. For the opaque, diffuse emitting and reflecting wall, the inflow radiative intensity ($\mathbf{n}_w \bullet \mathbf{\Omega} < 0$) is given as

$$I_0(\mathbf{x}_w,\mathbf{\Omega}) = \varepsilon_w I_b(\mathbf{x}_w) + \frac{1-\varepsilon_w}{\pi}\int_{\mathbf{n}_w \bullet \mathbf{\Omega}'>0} I(\mathbf{x}_w,\mathbf{\Omega}')\ \mathbf{\Omega}' \bullet \mathbf{n}_w \mathrm{d}\Omega' \tag{2}$$

where $\varepsilon_w$ is the wall emissivity.

In the following, the LSFEM discretization of the RTE is presented. In the LSFEM, the solid angular space discretization is by the common discrete ordinates approach. The radiative intensity for each discrete direction $\mathbf{\Omega}_m$ can be approximated using the FEM shape functions $\phi_i$ at each solution nodes, namely

$$\tilde{I}_m(\mathbf{x}) \simeq \sum_{i=1}^{N_{sol}} I_{m,i}\phi_i(\mathbf{x}) \tag{3}$$

where $I_{m,i}$ are the expansion coefficients, which also hold the values of the radiative intensity of direction $\mathbf{\Omega}_m$ at node *i* (namely, $I_{m,i} = I\left(\mathbf{\Omega}_m, \mathbf{x}_i\right)$) due to the Kronecker delta property of the FEM shape functions.

The theoretical advantage of the LSFEM discretization is that it is derived based on the minimization of the residual norms $\|R_m\| = \|L[\tilde{I}_m] - S_m\|$ of each direction. Here a short derivation is presented. The norm is defined in the solution domain as

$$\|R_m\| = \sqrt{< R_m, R_m >} \tag{4}$$

where the inner product $<\bullet,\bullet>$ is defined as

$$< f, g >= \int_V fg \mathrm{d}V \tag{5}$$

It is easily seen that the minimization of the norm $\|R_m\|$ is same as the minimization of the inner product $< R_m, R_m >$. By applying the FEM shape functions approximation, the minimized residual is obtained with the satisfaction of the following condition.

$$\frac{\partial < R_m, R_m >}{\partial I_{m,i}} = 0, \qquad i = 1, \ldots, N_{sol} \tag{6}$$

which further yields

$$< L[\tilde{I}_m] - S_m, L[\phi_i] >= 0, \qquad i = 1, \ldots, N_{sol} \tag{7}$$

Equation (7) can also be considered as a special weighted residual approach by taking the weighted function $W$ as $L[\phi_i] = \mathbf{\Omega}_m \bullet \nabla \phi_i + \beta \phi_i$. The LSFEM discretization of the RTE for each direction $\mathbf{\Omega}_m$ is then obtained as

$$\begin{aligned} &< \mathbf{\Omega}_m \bullet \nabla I, \mathbf{\Omega}_m \bullet \nabla \phi_j > + < \beta I, \beta \phi_j > + < \beta I, \mathbf{\Omega}_m \bullet \nabla \phi_j > + < \mathbf{\Omega}_m \bullet \nabla I, \beta \phi_j > \\ &= \ < S_m, \beta \phi_j > + < S_m, \mathbf{\Omega}_m \bullet \nabla \phi_j > \end{aligned} \tag{8}$$

which can be written in matrix form as

$$\mathbf{K}_m \mathbf{u}_m = \mathbf{h}_m \tag{9}$$

where the matrices are defined as

$$\mathbf{u}_m = \left[ u_{m,i} \right]_{i=1,N_{sol}} = \left[ I_m(\mathbf{x}_i) \right]_{i=1,N_{sol}} \tag{10a}$$

$$\begin{aligned} \mathbf{K}_m &= \left[ K_{m,ji} \right]_{j=1,N_{sol};i=1,N_{sol}} \\ &=< \mathbf{\Omega}_m \bullet \nabla \phi_i, \mathbf{\Omega}_m \bullet \nabla \phi_j > + < \beta \phi_i, \beta \phi_j > \\ &+ < \beta \phi_i, \mathbf{\Omega}_m \bullet \nabla \phi_j > + < \mathbf{\Omega}_m \bullet \nabla \phi_i, \beta \phi_j > \end{aligned} \tag{10b}$$

$$\mathbf{h}_m = \left[ h_{m,j} \right]_{j=1,N_{sol}} =< S_m, \beta \phi_j > + < S_m, \mathbf{\Omega}_m \bullet \nabla \phi_j > \tag{10c}$$

For two-dimensional media, the matrices can be further expanded as

$$\mathbf{K}_m = \mu_m^2 \mathbf{D}_{xx} + \mu_m \eta_m \mathbf{D}_{xy} + \mu_m \eta_m \left(\mathbf{D}_{xy}\right)^T + \eta_m^2 \mathbf{D}_{yy} + \left(\mu_m \mathbf{D}_x + \eta_m \mathbf{D}_y\right)\mathbf{B} + \mathbf{B}\left(\mu_m \mathbf{D}_x + \eta_m \mathbf{D}_y\right)^T + \mathbf{B}^2\mathbf{M} \tag{11a}$$

$$\mathbf{h}_m = \left[\mathbf{BM} + \left(\mu_m \mathbf{D}_x + \eta_m \mathbf{D}_y\right)\right]\mathbf{s}_m \tag{11b}$$

where

$$\mathbf{B} = diag\{\mathbf{b}\} \tag{12a}$$

$$\mathbf{b} = \left[b_i\right] = \left[\beta(\mathbf{x}_i)\right] \tag{12b}$$

$$\mathbf{s}_m = \left[s_{m,i}\right] = \left[S(\mathbf{x}_i, \mathbf{\Omega}_m)\right] \tag{12c}$$

$$\mathbf{D}_x = \left[D_{x,ji}\right] = \left[< \phi_i, \partial\phi_j / \partial x >\right] \tag{12d}$$

$$\mathbf{D}_y = \left[D_{y,ji}\right] = \left[< \phi_i, \partial\phi_j / \partial y >\right] \tag{12e}$$

$$\mathbf{D}_{xx} = \left[D_{xx,ji}\right] = \left[< \partial\phi_i / \partial x, \partial\phi_j / \partial x >\right] \tag{12f}$$

$$\mathbf{D}_{xy} = \left[D_{xy,ji}\right] = \left[< \partial\phi_i / \partial y, \partial\phi_j / \partial x >\right] \tag{12g}$$

$$\mathbf{D}_{yy} = \left[D_{yy,ji}\right] = \left[< \partial\phi_i / \partial y, \partial\phi_j / \partial y >\right] \tag{12h}$$

$$\mathbf{M} = \left[M_{ji}\right] = \left[< \phi_i, \phi_j >\right] \tag{12i}$$

For one-dimensional media, the matrices $\mathbf{K}_m$ and $\mathbf{h}_m$ can be further expanded as

$$\mathbf{K}_m = \xi_m^2 \mathbf{D}_{xx} + \xi_m \left(\mathbf{D}_x \mathbf{B} + \mathbf{B}\,\mathbf{D}_x^{\,T}\right) + \mathbf{B}^2\mathbf{M} \tag{13a}$$

$$\mathbf{h}_m = \left[\mathbf{BM} + \xi_m \mathbf{D}_x\right]\mathbf{s}_m \tag{13b}$$

where

$$\mathbf{D}_x = \left[D_{x,ji}\right] = \left[< \phi_i, \mathrm{d}\phi_j / \mathrm{d}x >\right] \tag{14a}$$

$$\mathbf{D}_{xx} = \left[D_{xx,ji}\right] = \left[< \mathrm{d}\phi_i / \mathrm{d}x, \mathrm{d}\phi_j / \mathrm{d}x >\right] \tag{14b}$$

It is noted that the following approximated relation is used in the formulation of stiff matrices from Eq. (10) to Eq. (11) and Eq.(13).

$$\phi_i(\mathbf{x})\, f(\mathbf{x}) \approx \phi_i(\mathbf{x})\, f_i \tag{15}$$

This approximation is established based on the compact support property of the FEM shape function.

As for the GFEM, the discretization can also be written in matrix form as Eq (9), and the stiff matrices can be obtained for two-dimensional media as

$$\mathbf{K}_m = \mu_m \mathbf{D}_x + \eta_m \mathbf{D}_y + \mathbf{MB} \tag{16a}$$

$$\mathbf{h}_m = \mathbf{M}\, \mathbf{s}_m \tag{16b}$$

and for one-dimensional media as

$$\mathbf{K}_m = \xi_m \mathbf{D}_x + \mathbf{MB} \tag{17a}$$

$$\mathbf{h}_m = \mathbf{M}\, \mathbf{s}_m \tag{17b}$$

*2.2. Boundary condition treatment and implementation*

The boundary condition (Eq. (2)) for the RTE is of Dirichlet type and often called the essential boundary condition. The accuracy in the imposing of this type of boundary condition is very important for the overall solution accuracy. One of the most accurate imposing of this kind of boundary condition is by operator collocation approach. In this approach, the row of stiff matrix $\mathbf{K}_m$ corresponding to the inflow boundary nodes is replaced with the discrete operator of the related boundary condition. Similar modification are also applied to the load vector $\mathbf{h}_m$. The modification process of the stiff matrix $\mathbf{K}_m$ and load vector $\mathbf{h}_m$ are formulated as follows (the modification is only conducted for nodes on the inflow boundary, namely, $\mathbf{n}_w(\mathbf{x}_j) \bullet \mathbf{\Omega}_m < 0$).

$$K_{m,ji} = \delta_{ji}\,, \tag{18a}$$

$$h_{m,j} = I_0(\mathbf{x}_j, \mathbf{\Omega}_m)\,. \tag{18b}$$

The solution procedures of the LSFEM are similar to other discrete ordinates based methods, which comprise an outer iteration of global source update and an inner angular loop.

## 3. Theoretical Analysis of the Stability and the Accuracy of the LSFEM

Though currently the LSFEM has been successfully applied to solving radiative transfer in homogeneous media with limited numerical verifications, the accuracy and stability of the LSFEM have not been well investigated theoretically. As will be demonstrated in this section, the LSFEM will suffer a deficiency problem for radiative transfer in strongly inhomogeneous media. This violates the traditional impression on the performance of the LSFEM in dealing with radiative transfer in homogeneous media and has not been noticed before.

In this section, the stability and the accuracy of the LSFEM are theoretically analyzed for both homogeneous and strongly inhomogeneous media. The focus is on the deficiency problem of the LSFEM for solving radiative transfer in strongly inhomogeneous media. Without loss of generality and for the convenience of theoretical analysis, the one-dimensional LSFEM formulation and the linear element is considered in the FEM discretization on a uniform grid.

*3.1. Explicit formulation of the LSFEM and its equivalence to the LSORTE*

To proceed the theoretical analysis, it is necessary to obtain the explicit formulation of the LSFEM at a specific node. For the 1D linear element, the global shape function at node *i* can be written as

$$\phi_i(x) = \begin{cases} (x - x_{i-1}) / \Delta x, & x_{i-1} \le x < x_i \\ (x_{i+1} - x) / \Delta x, & x_i \le x < x_{i+1} \\ 0, & \text{otherwise} \end{cases}, \tag{19}$$

which is also graphically shown in Figure 1 for better understanding.

Because the shape function $\phi_i(x)$ is only nonzero in the two connecting elements around node *i*, the elements of the stiff matrix $\mathbf{K}_m$ may have only nonzero values for entries those related to the neighboring nodes. This is similar to the finite difference method. Referring the stiff matrix given in Eq. (13), the *i*-th row of the stiff matrix $\mathbf{K}_m$ can be written explicitly with the knowledge of the following integrations.

$$<\phi_i, \phi_{i+1}> = <\phi_i, \phi_{i-1}> = \int_0^{\Delta x} \left(1 - \frac{x}{\Delta x}\right) \frac{x}{\Delta x} \mathrm{d}x = \frac{1}{6} \Delta x \tag{20a}$$

$$<\phi_i, \phi_i> = 2 \int_0^{\Delta x} \left(1 - \frac{x}{\Delta x}\right)\left(1 - \frac{x}{\Delta x}\right) \mathrm{d}x = \frac{2}{3} \Delta x \tag{20b}$$

$$<\frac{\mathrm{d}}{\mathrm{d}x}\phi_i, \phi_{i+1}> = \int_0^{\Delta x} \frac{\mathrm{d}}{\mathrm{d}x}\left[\left(1 - \frac{x}{\Delta x}\right)\right] \frac{x}{\Delta x} \mathrm{d}x = -\frac{1}{2} \tag{20c}$$

$$<\frac{\mathrm{d}}{\mathrm{d}x}\phi_i,\phi_{i-1}>=\int_{-\Delta x}^{0}\frac{\mathrm{d}}{\mathrm{d}x}\left[\left(1+\frac{x}{\Delta x}\right)\right]\frac{-x}{\Delta x}\mathrm{d}x=\frac{1}{2} \tag{20d}$$

$$<\phi_i,\frac{\mathrm{d}}{\mathrm{d}x}\phi_{i+1}>=\int_{0}^{\Delta x}\left(1-\frac{x}{\Delta x}\right)\frac{\mathrm{d}}{\mathrm{d}x}\left[\frac{x}{\Delta x}\right]\mathrm{d}x=\frac{1}{2} \tag{20e}$$

$$<\phi_i,\frac{\mathrm{d}}{\mathrm{d}x}\phi_{i-1}>=\int_{-\Delta x}^{0}\left(1+\frac{x}{\Delta x}\right)\frac{\mathrm{d}}{\mathrm{d}x}\left[\frac{-x}{\Delta x}\right]\mathrm{d}x=-\frac{1}{2} \tag{20f}$$

$$\begin{aligned}<\phi_i,\frac{\mathrm{d}}{\mathrm{d}x}\phi_i>&=\int_{-\Delta x}^{0}\left(1+\frac{x}{\Delta x}\right)\frac{\mathrm{d}}{\mathrm{d}x}\left[\left(1+\frac{x}{\Delta x}\right)\right]\mathrm{d}x\\&+\int_{0}^{\Delta x}\left(1-\frac{x}{\Delta x}\right)\frac{\mathrm{d}}{\mathrm{d}x}\left[\left(1-\frac{x}{\Delta x}\right)\right]\mathrm{d}x=0\end{aligned} \tag{20g}$$

$$<\frac{\mathrm{d}}{\mathrm{d}x}\phi_i,\frac{\mathrm{d}}{\mathrm{d}x}\phi_{i+1}>=\int_{0}^{\Delta x}\frac{\mathrm{d}}{\mathrm{d}x}\left[\left(1-\frac{x}{\Delta x}\right)\right]\frac{\mathrm{d}}{\mathrm{d}x}\left[\frac{x}{\Delta x}\right]\mathrm{d}x=-\frac{1}{\Delta x} \tag{20h}$$

$$<\frac{\mathrm{d}}{\mathrm{d}x}\phi_i,\frac{\mathrm{d}}{\mathrm{d}x}\phi_{i-1}>=\int_{-\Delta x}^{0}\frac{\mathrm{d}}{\mathrm{d}x}\left[\left(1+\frac{x}{\Delta x}\right)\right]\frac{\mathrm{d}}{\mathrm{d}x}\left[\frac{-x}{\Delta x}\right]\mathrm{d}x=-\frac{1}{\Delta x} \tag{20i}$$

$$\begin{aligned}<\frac{\mathrm{d}}{\mathrm{d}x}\phi_i,\frac{\mathrm{d}}{\mathrm{d}x}\phi_i>&=\int_{-\Delta x}^{0}\frac{\mathrm{d}}{\mathrm{d}x}\left[\left(1+\frac{x}{\Delta x}\right)\right]\frac{\mathrm{d}}{\mathrm{d}x}\left[\left(1+\frac{x}{\Delta x}\right)\right]\mathrm{d}x\\&+\int_{0}^{\Delta x}\frac{\mathrm{d}}{\mathrm{d}x}\left[\left(1-\frac{x}{\Delta x}\right)\right]\frac{\mathrm{d}}{\mathrm{d}x}\left[\left(1-\frac{x}{\Delta x}\right)\right]\mathrm{d}x=\frac{2}{\Delta x}\end{aligned} \tag{20j}$$

It is noted the integration domain is the support domain of $\phi_i(x)$ which is centered at $x=x_i$ and is translated to the origin to ease the calculation. By substitution of the integrations Eqs. (20) into Eqs. (13), the explicit formulation of the LSFEM at node *i* is obtained as

$$\begin{aligned}&\xi_m^2\begin{bmatrix}-1/\Delta x\\2/\Delta x\\-1/\Delta x\end{bmatrix}^T\begin{bmatrix}I_{m,i-1}\\I_{m,i}\\I_{m,i+1}\end{bmatrix}+\xi_m\begin{bmatrix}1/2\\0\\-1/2\end{bmatrix}^T\begin{bmatrix}\beta_{i-1}&&\\&\beta_i&\\&&\beta_{i+1}\end{bmatrix}\begin{bmatrix}I_{m,i-1}\\I_{m,i}\\I_{m,i+1}\end{bmatrix}\\&+\xi_m\beta_i\begin{bmatrix}-1/2\\0\\1/2\end{bmatrix}^T\begin{bmatrix}I_{m,i-1}\\I_{m,i}\\I_{m,i+1}\end{bmatrix}+\beta_i^2\begin{bmatrix}\Delta x/6\\2\Delta x/3\\\Delta x/6\end{bmatrix}^T\begin{bmatrix}I_{m,i-1}\\I_{m,i}\\I_{m,i+1}\end{bmatrix}\\&=\beta_i\begin{bmatrix}\Delta x/6\\2\Delta x/3\\\Delta x/6\end{bmatrix}^T\begin{bmatrix}S_{m,i-1}\\S_i\\S_{m,i+1}\end{bmatrix}+\xi_m\begin{bmatrix}1/2\\0\\-1/2\end{bmatrix}^T\begin{bmatrix}S_{m,i-1}\\S_{m,i}\\S_{m,i+1}\end{bmatrix}\end{aligned} \tag{21}$$

By dividing $\Delta x$, Eq. (21) can also be written in an equivalence form as

$$
\begin{aligned}
&\xi_m^2 \frac{1}{(\Delta x)^2}\begin{bmatrix}-1\\2\\-1\end{bmatrix}^T\begin{bmatrix}I_{m,i-1}\\I_{m,i}\\I_{m,i+1}\end{bmatrix}+\xi_m\frac{1}{\Delta x}\begin{bmatrix}1/2\\0\\-1/2\end{bmatrix}^T\begin{bmatrix}\beta_{i-1}I_{m,i-1}\\\beta_i I_{m,i}\\\beta_{i+1}I_{m,i+1}\end{bmatrix}\\
&+\xi_m\beta_i\frac{1}{\Delta x}\begin{bmatrix}-1/2\\0\\1/2\end{bmatrix}^T\begin{bmatrix}I_{m,i-1}\\I_{m,i}\\I_{m,i+1}\end{bmatrix}+\beta_i^2\begin{bmatrix}1/6\\2/3\\1/6\end{bmatrix}^T\begin{bmatrix}I_{m,i-1}\\I_{m,i}\\I_{m,i+1}\end{bmatrix}\\
&=\beta_i\begin{bmatrix}1/6\\2/3\\1/6\end{bmatrix}^T\begin{bmatrix}S_{m,i-1}\\S_{m,i}\\S_{m,i+1}\end{bmatrix}+\xi_m\frac{1}{\Delta x}\begin{bmatrix}1/2\\0\\-1/2\end{bmatrix}^T\begin{bmatrix}S_{m,i-1}\\S_{m,i}\\S_{m,i+1}\end{bmatrix}
\end{aligned}
\tag{22}
$$

The theoretical analysis presented in the next section is based on this explicit form of the LSFEM (Eq. (22)).

As a reference, the one-dimensional explicit formulation of the GFEM can be written following the similar procedure as

$$
\xi_m\frac{1}{\Delta x}\begin{bmatrix}-1/2\\0\\1/2\end{bmatrix}^T\begin{bmatrix}I_{m,i-1}\\I_{m,i}\\I_{m,i+1}\end{bmatrix}+\begin{bmatrix}1/6\\2/3\\1/6\end{bmatrix}^T\begin{bmatrix}\beta_{i-1}I_{m,i-1}\\\beta_i I_{m,i}\\\beta_{i+1}I_{m,i+1}\end{bmatrix}=\begin{bmatrix}1/6\\2/3\\1/6\end{bmatrix}^T\begin{bmatrix}S_{m,i-1}\\S_{m,i}\\S_{m,i+1}\end{bmatrix}
\tag{23}
$$

To get further understanding of the LSFEM, here we show that the LSFEM discretization is equivalent to a central difference discretization of a kind of second order radiative transfer equation, which has just been proposed by the same author [35] and named as the least squares second order radiative transfer equation (LSORTE). By discretization with central difference like numerical method, such as the FEM and the meshless methods, the second order forms of radiative transfer equation shows better numerical properties than the RTE. Besides the even parity formulation of the RTE, there are three second order forms of the radiative transfer equations have been proposed, the SORTE [27, 36], the LSORTE [35] and the MSORTE [35]. The performance of the second order radiative transfer equations have been studied by the author in several papers [35].

It is easily shown that the discrete operator $\frac{1}{(\Delta x)^2}\begin{bmatrix}-1 & 2 & -1\end{bmatrix}$ is exactly the central difference discretization of the second order differential operator of $-\frac{\mathrm{d}^2}{\mathrm{d}x^2}$, and $\frac{1}{\Delta x}\begin{bmatrix}-1/2 & 0 & 1/2\end{bmatrix}$ is exactly the central difference discretization of the first order differential operator of $\frac{\mathrm{d}}{\mathrm{d}x}$. Furthermore, by using the Taylor expansion to expand the $I_{m,i-1}$ and $I_{m,i+1}$ at node *i*, it is easy shown that

$$\begin{bmatrix} 1/6 \\ 2/3 \\ 1/6 \end{bmatrix}^T \begin{bmatrix} I_{m,i-1} \\ I_{m,i} \\ I_{m,i+1} \end{bmatrix} = I_{m,i} + O(\Delta x^2) \tag{24}$$

Hence the discrete operator $\begin{bmatrix} 1/6 & 2/3 & 1/6 \end{bmatrix}$ can be considered as a discretization of the identity operator with second order accuracy.

As a result, by substitution with these corresponding differential operator, Eq. (22) is equivalent to the following differential equation.

$$\begin{aligned} &-\xi_m^2 \frac{\mathrm{d}^2 I}{\mathrm{d}x^2} - \xi_m \frac{\mathrm{d}\beta I}{\mathrm{d}x} + \xi_m \beta \frac{\mathrm{d}I}{\mathrm{d}x} + \beta^2 I \\ &= \beta S - \xi_m \frac{\mathrm{d}S}{\mathrm{d}x} \end{aligned} \tag{25}$$

which can be written in ray path coordinate $s$ as

$$\begin{aligned} &-\frac{\mathrm{d}^2 I}{\mathrm{d}s^2} - \frac{\mathrm{d}\beta I}{\mathrm{d}s} + \beta \frac{\mathrm{d}I}{\mathrm{d}s} + \beta^2 I \\ &= \beta S - \frac{\mathrm{d}S}{\mathrm{d}s} \end{aligned} \tag{26}$$

and is equivalent to

$$-\frac{\mathrm{d}^2 I}{\mathrm{d}s^2} + \left( \beta^2 - \frac{\mathrm{d}\beta}{\mathrm{d}s} \right) I = \beta S - \frac{\mathrm{d}S}{\mathrm{d}s} \tag{27}$$

This is just the LSORTE [35] in ray path coordinate. Hence it is demonstrated that the LSFEM is equivalent to the LSORTE with central difference discretization.

*3.2. Stability and accuracy for homogeneous media*

It is difficult to analyze the case of variable extinction coefficient due to the theoretical difficulty in dealing with variable coefficient differential equations. Here we begin with the case of homogenous media. For homogenous media at direction $\xi_m = 1$, the explicit form of the LSFEM (Eq. (22)) can be simplified as

$$
\begin{aligned}
&\left\{\frac{1}{(\Delta x)^2}\begin{bmatrix}-1\\2\\-1\end{bmatrix}^T+\beta\frac{1}{\Delta x}\begin{bmatrix}1/2\\0\\-1/2\end{bmatrix}^T+\beta\frac{1}{\Delta x}\begin{bmatrix}-1/2\\0\\1/2\end{bmatrix}^T+\beta^2\begin{bmatrix}1/6\\2/3\\1/6\end{bmatrix}^T\right\}\begin{bmatrix}I_{i-1}\\I_i\\I_{i+1}\end{bmatrix}\\
&=\left(\beta\begin{bmatrix}1/6\\2/3\\1/6\end{bmatrix}^T+\frac{1}{\Delta x}\begin{bmatrix}1/2\\0\\-1/2\end{bmatrix}^T\right)\begin{bmatrix}S_{i-1}\\S_i\\S_{i+1}\end{bmatrix}
\end{aligned}
\tag{28}
$$

Though an analysis on any direction can be obtained, the selection of direction $\xi_m = 1$ is for brevity and without loss of generality. The discrete equation (Eq. (28)) at node *i* can be further written in its equivalent differential form at node *i* by using Taylor expansions to the values of intensity and source term at other nodes. For example,

$$
I_{i+1} = I_i + \Delta x\frac{\partial I_i}{\partial x} + \sum_{k=2}^{\infty}(\Delta x)^k\frac{1}{k!}\frac{\partial^k I_i}{\partial x^k} \tag{29a}
$$

$$
I_{i-1} = I_i - \Delta x\frac{\partial I_i}{\partial x} + \sum_{k=2}^{\infty}(-1)^k\Delta x^k\frac{1}{k!}\frac{\partial^k I_i}{\partial x^k} \tag{29b}
$$

The substitution of these expansion Eqs. (29) into the LSFEM discretization (Eq. (28)) will yield the equivalent differential form at node *i*. With written in the differential form, the Fourier transform can then be applied to facilitate the analysis. Fourier transforms applied to these expansion yield

$$
\widehat{I_{i+1}} = \sum_{k=0}^{\infty}\frac{1}{k!}(j\varpi\Delta x)^k\,\widehat{I}_i = e^{j\varpi\Delta x}\,\widehat{I}_i \tag{30}
$$

$$
\widehat{I_{i-1}} = \sum_{k=0}^{\infty}\frac{1}{k!}(-j\varpi\Delta x)^k\,\widehat{I}_i = e^{-j\varpi\Delta x}\,\widehat{I}_i \tag{31}
$$

Here $j=\sqrt{-1}$ and $\varpi$ is the corresponding frequency domain variable. Similar relations can also be obtained for the source term, namely, $\widehat{S_{i+1}} = e^{j\varpi\Delta x}\widehat{S}_i$ and $\widehat{S_{i-1}} = e^{-j\varpi\Delta x}\widehat{S}_i$ . As such the exact form of the Eq. (28) in frequency domain can be written as

$$
\begin{aligned}
&\left\{\frac{1}{(\Delta x)^2}\begin{bmatrix}-1\\2\\-1\end{bmatrix}^T+\beta\frac{1}{\Delta x}\begin{bmatrix}1/2\\0\\-1/2\end{bmatrix}^T\right.\\
&\left.+\beta\frac{1}{\Delta x}\begin{bmatrix}-1/2\\0\\1/2\end{bmatrix}^T+\beta^2\begin{bmatrix}1/6\\2/3\\1/6\end{bmatrix}^T\right\}\begin{bmatrix}e^{-j\varpi\Delta x}\\1\\e^{j\varpi\Delta x}\end{bmatrix}\widehat{I}_i\\
&=\left(\beta\begin{bmatrix}1/6\\2/3\\1/6\end{bmatrix}^T+\frac{1}{\Delta x}\begin{bmatrix}1/2\\0\\-1/2\end{bmatrix}^T\right)\begin{bmatrix}e^{-j\varpi\Delta x}\\1\\e^{j\varpi\Delta x}\end{bmatrix}\widehat{S}_i
\end{aligned} \tag{32}
$$

which can be further written as

$$
\begin{aligned}
&\left[\varpi^2\mathrm{sinc}^2\left(\frac{\varpi\Delta x}{2}\right)+\beta^2\left(\frac{1}{3}\cos(\varpi\Delta x)+\frac{2}{3}\right)\right]\widehat{I}_i\\
&=\left[\beta\left(\frac{1}{3}\cos(\varpi\Delta x)+\frac{2}{3}\right)-j\varpi\mathrm{sinc}(\varpi\Delta x)\right]\widehat{S}_i
\end{aligned} \tag{33}
$$

It is thus obtained that

$$
\widehat{I}_i=\frac{\beta\left(\frac{1}{3}\cos(\varpi\Delta x)+\frac{2}{3}\right)-j\varpi\mathrm{sinc}(\varpi\Delta x)}{\varpi^2\mathrm{sinc}^2\left(\frac{\varpi\Delta x}{2}\right)+\beta^2\left(\frac{1}{3}\cos(\varpi\Delta x)+\frac{2}{3}\right)}\widehat{S}_i \tag{34}
$$

It is noted that the exact intensity $I_E$ at node *i* obey the RTE taking the form as

$$
\frac{\mathrm{d}I_E}{\mathrm{d}x}+\beta I_E=S \tag{35}
$$

And it is obtained from frequency domain analysis that

$$
\widehat{I_E}=\frac{1}{j\varpi+\beta}\hat{S} \tag{36}
$$

As such, a relative error for the LSFEM discretization in frequency domain can be defined and obtained as

$$
\begin{aligned}
E_{LSFEM}&=\frac{\hat{I}_i-\hat{I}_E}{\hat{I}_E}\\
&=\frac{\left[\left(\frac{1}{3}\cos(2\pi\bar{\varpi})+\frac{2}{3}\right)-j2\pi\bar{\varpi}\tau_\Delta^{-1}\mathrm{sinc}(2\pi\bar{\varpi})\right]\left(j2\pi\bar{\varpi}\tau_\Delta^{-1}+1\right)}{\left(2\pi\bar{\varpi}\tau_\Delta^{-1}\right)^2\mathrm{sinc}^2(\pi\bar{\varpi})+\left(\frac{1}{3}\cos(2\pi\bar{\varpi})+\frac{2}{3}\right)}-1
\end{aligned} \tag{37}
$$

in which, $\varpi$ is a reduced frequency defined as $\varpi = \dfrac{\Delta x \varpi}{2\pi}$ and $\tau_\Delta = \beta \Delta x$ is the grid optical thickness.

As for the GFEM discretization (Eq. (23)), the exact form at direction $\mu_m = 1$ in frequency domain can be obtained as

$$\left[ j\varpi \operatorname{sinc}(\varpi \Delta x) + \beta \left( \frac{1}{3}\cos(\varpi \Delta x) + \frac{2}{3} \right) \right] \widehat{I}_i = \left( \frac{1}{3}\cos(\varpi \Delta x) + \frac{2}{3} \right) \widehat{S}_i \tag{38}$$

Similarly, a relative error for the GFEM discretization in frequency domain can be defined and obtained as

$$\begin{aligned} E_{GFEM} &= \frac{\widehat{I}_i - \widehat{I}_E}{\widehat{I}_E} = \frac{\widehat{I}_i}{\widehat{I}_E} - 1 \\ &= \frac{j2\pi\varpi\tau_\Delta^{-1} + 1}{j2\pi\varpi\tau_\Delta^{-1}\operatorname{sinc}(2\pi\varpi) + \left( \frac{1}{3}\cos(2\pi\varpi) + \frac{2}{3} \right)} - 1 \end{aligned} \tag{39}$$

A comparison of the solution error of the LSFEM and the GFEM at different grid optical thickness is presented in Figure 2. The frequency range of the reduced frequency $\varpi$ is plotted in [0, 0.5]. This is based on the fact that the maximum frequency (or shortest wavelength) of a harmonic that can propagate on a uniform grid of spacing $\Delta s$ is $\pi / \Delta s$ (or wavelength $2\Delta s$), namely, $\varpi$ = 0.5. For both the GFEM and the LSFEM, it is seen that the relative error increases with $\varpi$ for different grid optical thickness, and the maximum relative error occur at $\varpi$ = 0.5. However, at high frequency, e.g. $\varpi$ = 0.5, the relative error of the GFEM is more than two orders of magnitude greater than that of the LSFEM at grid optical thickness of $\tau_\Delta$=0.01. As a result, significant high frequency error (known as spurious ‘wiggles’) can be observed in the spatial domain solution even if the exact solution has a relatively small component at the high frequency near $\varpi$=0.5. That is because the solution errors of that component will be magnified to several orders of magnitude. Referring Eq. (36), it indicates that if the source function $S$ has high frequency component, then the exact solution will contain high frequency component. This is true for the case in which the emission field contains large gradient because the large gradient will induce a wide band spectrum distribution. One such example is the Gaussian emission field model problem introduced in Ref. [27].

Furthermore, it is noticed from Figure 2 that the high frequency error of the GFEM tends to decrease with the increase of the grid optical thickness. A detailed relation of high frequency error (for $\varpi = 0.5$) with grid optical thickness for the GFEM and the LSFEM are presented in Figure 3. The performance of the LSFEM is superior to the GFEM at high frequency error for a wide range of grid optical thickness. This proves and confirms the good stability of the LSFEM.

These theoretical results interpret the stability problem of the GFEM [27] in dealing with such kind of problems well, namely, 1) the strong spurious wiggles appears in the solution, 2) the spurious wiggles has a short wavelength about double of the grid spacing and 3) the magnitude of the spurious wiggles tends to decrease with the increasing of extinction coefficient.

### *3.3. Accuracy in dealing with strongly inhomogeneous medium*

When the LSFEM is applied to solve radiative transfer in strongly inhomogeneous media, a deficiency problem is observed. The LSFEM will show a bad accuracy for solving radiative transfer in strongly inhomogeneous media. This phenomena has not been yet studied before. This section intends to analytically interpret this problem and evaluate the performance of the LSFEM for strongly inhomogeneous media.

The deficiency problem of the LSFEM for solving radiative transfer in strongly inhomogeneous media is demonstrated by solving the following model problem, namely, the radiative transfer in a purely absorbing inhomogeneous infinite slab. The absorption coefficient is a discontinuous function of position given as

$$\kappa_a(x) = \bar{\kappa}_a U\left(x - 0.5L\right), \quad x \in [0, L] \tag{40}$$

where $\bar{\kappa}_a$ is a constant parameter indicating the strength of extinction, $L$ is the slab thickness and $U(\bullet)$ is the unit step function. Both of the walls are black, and the medium is nonemitting. The intensity distribution $I(x,\xi)$ for direction $\xi = 1$ and $L = 10$ is solved by using the LSFEM and the GFEM. The boundary intensity is prescribed as unity. The exact analytical solution of this problem is

$$I_E^{\text{mo}}(x) = U(x - 0.5L)e^{-\bar{\kappa}_a(x-0.5L)} + U(0.5L - x) \tag{41}$$

This problem was first introduced in Ref. [35] to study the numerical performance of the LSORTE.

The solved intensity distribution by the GFEM and the LSFEM are presented in Figure 4 for different $\bar{\kappa}_a$ and are compared with the exact solutions. The slab is subdivided uniformly into 100 linear elements. As for the results obtained using the GFEM [Figure 4(a)], two facts can be observed: (1) obvious spurious wiggles (or high frequency error) exist in the results, and (2) the results generally agree well with the exact solutions. These observations seem agree with the theoretical prediction for homogeneous media presented in Section 3.2. As for the results obtained using the LSFEM [Figure 4(b)], two facts emphasized are (1) no spurious wiggles appear in the results, and (2) particularly large errors are observed in the results and the errors increase significantly with the increase of $\bar{\kappa}_a$. The first fact is attributed to the good stability of the LSFEM which agrees with the traditional understanding of method and the theoretical prediction presented in Section 3.2 for homogeneous media. However, the second fact violates the good performance of the LSFEM as traditionally known and is inconsistent with the previous theoretical prediction of the performance of the LSFEM for homogeneous media. This is a strange deficiency problem of the LSFEM for solving radiative transfer in strongly inhomogeneous media and it has not been addressed before. As such, performance of numerical methods in solving of radiative transfer in strongly inhomogeneous media needs to be checked carefully. Theoretically results obtained based on the analysis of numerical methods for solving homogeneous media cannot be taken directly for strongly inhomogeneous media without thorough verification.

In the following, a theoretical analysis in frequency domain is conducted to interpret the deficiency problem of the LSFEM. Because the extinction coefficient is variable, the Fourier analysis presented in Section 3.2 can not be directly applied to analyze the accuracy of the LSFEM for solving radiative transfer in strongly inhomogeneous media. Here, a special source-linearization technique proposed in Ref. [35] is used to facilitate the using of Fourier analysis, which transforms the original variable coefficient equation into an equivalent constant coefficient equation operable by Fourier transform. The technique is based on the equivalence of the solutions to the RTE with the following equation under the same boundary conditions.

$$\boldsymbol{\Omega}\bullet\nabla I = S - \beta I_E \tag{42}$$

in which $I_E$ is the exact solution of the RTE under the definite boundary condition. For one dimensional media, the source-linearization technique applied to the discrete form of the LSFEM result in the following modified stiff matrix and load vector.

$$\mathbf{K}_m = \xi_m^2 \mathbf{D}_{xx} \tag{43a}$$

$$\mathbf{h}_m = \left[\mathbf{BM} + \xi_m \mathbf{D}_x\right]\mathbf{s}_m - \xi_m\left(\mathbf{D}_x\mathbf{B} + \mathbf{B}\,\mathbf{D}_x^{\ T}\right)\mathbf{t}_m - \mathbf{B}^2\mathbf{M}\,\mathbf{t}_m \tag{43b}$$

in which

$$\mathbf{t}_m = \left[t_{m,i}\right] = \left[I_E(\mathbf{x}_i, \boldsymbol{\Omega}_m)\right] \tag{44}$$

As for the GFEM with using the source-linearization technique, the resulting stiff matrix and load vector for one dimensional media are

$$\mathbf{K}_m = \xi_m \mathbf{D}_x \tag{45a}$$

$$\mathbf{h}_m = \mathbf{M}\,\mathbf{s}_m - \mathbf{M}\,\mathbf{B}\,\mathbf{t}_m \tag{45b}$$

It may be reasonable to conjecture that the standard LSFEM (presented in Section 2) will share some similar numerical property as the LSFEM using the source-linearization technique (stiff matrix defiend by Eq. (43)), such as both suffer from the deficiency problem. If this is true, then the analysis of the deficiency problem of the LSFEM in solving the RTE can be conducted by the analysis of the LSFEM using the source-linearization technique. This conjecture is not easily proved theoretically, however, it can be verified through numerical experiment on a model problem with the known exact solution.

A numerical experiment is conducted in the following. Numerical solutions of the previous mentioned model problem by the GFEM and the LSFEM with using source-linearization technique for different $\bar{\kappa}_a$ are presented in Figure 5 and Figure 6, respectively. The computational condition is the same as the results obtained without using the source-linearization. It is clearly observed that similar numerical behavior retains:

(1) the wiggles retains in the results of the GFEM with using source-linearization technique, (2) the deficiency problem retains for the LSFEM with using source-linearization technique, namely, particularly large errors are observed in the results and the errors increase significantly with the increase of $\bar{\kappa}_a$. Hence conjecture the mentioned above is proved numerically.

In this understanding, the analysis of the deficiency problem of the LSFEM can be conducted based on the analysis of the LSFEM with using the source-linearization technique. In the following, a frequency domain analysis of the accuracy of the GFEM and the LSFEM with using the source-linearization technique is conducted. Following the procedure presented in Section 3.1, the LSFEM discretization with using the source-linearization technique (Eq.(43)) for the model problem for $\xi = 1$ can be written explicitly at node $i$ as

$$\begin{aligned}\frac{1}{(\Delta x)^2}\begin{bmatrix}-1\\2\\-1\end{bmatrix}^T\begin{bmatrix}I_{i-1}\\I_i\\I_{i+1}\end{bmatrix}&=\frac{1}{\Delta x}\begin{bmatrix}-1/2\\0\\1/2\end{bmatrix}^T\begin{bmatrix}\kappa_{a;i-1}I_{E;i-1}^{\mathrm{mo}}\\\kappa_{a;i}I_{E;i}^{\mathrm{mo}}\\\kappa_{a;i+1}I_{E;i+1}^{\mathrm{mo}}\end{bmatrix}\\&-\kappa_{a;i}\frac{1}{\Delta x}\begin{bmatrix}-1/2\\0\\1/2\end{bmatrix}^T\begin{bmatrix}I_{E;i-1}^{\mathrm{mo}}\\I_{E;i}^{\mathrm{mo}}\\I_{E;i+1}^{\mathrm{mo}}\end{bmatrix}-\kappa_{a;i}^2\begin{bmatrix}1/6\\2/3\\1/6\end{bmatrix}^T\begin{bmatrix}I_{E;i-1}^{\mathrm{mo}}\\I_{E;i}^{\mathrm{mo}}\\I_{E;i+1}^{\mathrm{mo}}\end{bmatrix}\end{aligned} \tag{46}$$

and the GFEM discretization with using the source-linearization technique for the model problem discretization with using the source-linearization technique as

$$\frac{1}{\Delta x}\begin{bmatrix}-1/2\\0\\1/2\end{bmatrix}^T\begin{bmatrix}I_{m,i-1}\\I_{m,i}\\I_{m,i+1}\end{bmatrix}=-\begin{bmatrix}1/6\\2/3\\1/6\end{bmatrix}^T\begin{bmatrix}\kappa_{a;i-1}I_{E;i-1}^{\mathrm{mo}}\\\kappa_{a;i}I_{E;i}^{\mathrm{mo}}\\\kappa_{a;i+1}I_{E;i+1}^{\mathrm{mo}}\end{bmatrix} \tag{47}$$

respectively.

It is noted that both the LSFEM and the GFEM use the same scheme to imposing the same boundary condition. Hence the differences in their performance are due to the different discretizations in the medium. These differences in the performance can be investigated by the Fourier analysis given below.

Following the Fourier analysis presented in Section 3.2, Eq. (46) can be written in Frequency domain as

$$\varpi^2 \mathrm{sinc}^2\left(\frac{\varpi \Delta x}{2}\right)\widehat{I}_i = j\varpi \mathrm{sinc}\left(\varpi \Delta x\right)\widehat{\kappa_{a;i} I_{E;i}^{\mathrm{mo}}} - \widehat{F} \tag{48}$$

in which $\widehat{F}$ is defined as

$$\widehat{F} = \kappa_{a;i}\widehat{\frac{I_{E;i+1}^{\mathrm{mo}} - I_{E;i-1}^{\mathrm{mo}}}{2\Delta x}} + \kappa_{a;i}^2 \widehat{\left(\frac{1}{6} I_{E;i-1}^{\mathrm{mo}} + \frac{2}{3} I_{E;i}^{\mathrm{mo}} + \frac{1}{6} I_{E;i+1}^{\mathrm{mo}}\right)} \tag{49}$$

and Eq. (47) can be written in Frequency domain as

$$j\varpi \mathrm{sinc}\left(\varpi \Delta x\right)\widehat{I}_i = -\left(\frac{1}{3}\cos\left(\varpi \Delta x\right) + \frac{2}{3}\right)\widehat{\kappa_{a;i} I_{E;i}^{\mathrm{mo}}} \tag{50}$$

respectively.

A Fourier transform of Eq. (42) for the model problem yields $j\varpi \hat{I} = -\widehat{\kappa_a I_E^{\mathrm{mo}}}$, hence the relation for the exact solution in frequency domain can be obtained as

$$\hat{I}_E = -\frac{\widehat{\kappa_a I_E^{\mathrm{mo}}}}{j\varpi} \tag{51}$$

By using the frequency domain relations Eq. (48), Eq. (50) and (51), the relative error in frequency domain of the LSFEM and the GFEM discretization with using the source-linearization technique for the model problem can be defined and obtained as

$$\begin{aligned}
\tilde{E}_{LSFEM} &= \frac{\widehat{I}_i - \widehat{I}_E}{\widehat{I}_E} = \frac{\widehat{I}_i}{\widehat{I}_E} - 1 \\
&= \frac{-\left[j\varpi \mathrm{sinc}\left(\varpi \Delta x\right)\widehat{\kappa_{a;i} I_{E;i}^{\mathrm{mo}}} - \widehat{F}\right] j\varpi}{\varpi^2 \mathrm{sinc}^2\left(\varpi \Delta x/2\right)\widehat{\kappa_a I_E^{\mathrm{mo}}}} - 1 \\
&= \frac{\varpi \mathrm{sinc}\left(\varpi \Delta x\right) + j\widehat{F} / \widehat{\kappa_a I_E^{\mathrm{mo}}}}{\varpi \mathrm{sinc}^2\left(\varpi \Delta x/2\right)} - 1
\end{aligned} \tag{52}$$

and

$$\begin{aligned}
\tilde{E}_{GFEM} &= \frac{\widehat{I}_i}{\widehat{I}_E} - 1 = \frac{\frac{1}{3}\left(\cos(\varpi \Delta x) + 2\right)\widehat{\kappa_{a;i} I_{E;i}^{\mathrm{mo}}}\, j\varpi}{j\varpi \mathrm{sinc}\left(\varpi \Delta x\right)\widehat{\kappa_a I_E^{\mathrm{mo}}}} - 1 \\
&= \frac{\frac{1}{3}\left(\cos(\varpi \Delta x) + 2\right)}{\mathrm{sinc}\left(\varpi \Delta x\right)} - 1
\end{aligned} \tag{53}$$

respectively.

The modulus of these error relations with reduced frequency for three different value of grid optical thickness $\tau_{\Delta} = \bar{\kappa}_a \Delta x$ are plotted in Figure 7. The error relations of the LSORTE discretized by central difference scheme [35] and the MSORTE discretized by central difference scheme [35] for solving the model problem are also presented for references. It is noted that the Fourier transforms related to exact intensity, namely, $\widehat{\kappa_a I_E^{\mathrm{mo}}}$ and $\widehat{F}$ are calculated via the fast Fourier transform (FFT) and the functions are evaluated in the same spatial grid as that used in the numerical solution. The theoretical relations reveal the following important facts:

(1) The GFEM retains the large high frequency errors for solving radiative transfer in strongly inhomogeneous media as for solving radiative transfer in homogeneous media. This interprets the spurious wiggles in the numerical results of GFEM (Figure 5).

(2) The LSFEM retains the lower high frequency errors for solving radiative transfer in strongly inhomogeneous media. This interprets the stability of LSFEM, namely, the results are free of 'wiggles', for solving radiative transfer in strongly inhomogeneous media (Figure 6).

(3) The LSFEM shows very large low frequency errors for solving radiative transfer in strongly inhomogeneous media, and the low frequency error increase significantly with the increase of grid optical thickness $\tau_{\Delta} = \bar{\kappa}_a \Delta x$ or $\bar{\kappa}_a$. This interprets the deficiency problem of the LSFEM, namely, particularly large errors are observed in the results and the errors increase significantly with the increase of $\bar{\kappa}_a$. (shown in Figure 4(b)).

(4) The error distribution of the LSFEM is very similar to that of the LSORTE discretized by central difference scheme. This confirms the equivalence of the LSFEM discretization with the LSORTE discretized by central difference scheme (as discussed in Section 3.1 and in Ref. [35]).

Furthermore, the error distribution of the MSORTE discretized by central difference scheme is the only one which shows good performance both with low frequency error and high frequency error as compared to the LSFEM and the GFEM. This indicates that a FEM based on the MSORTE is promising as a cure of the deficiency problem of the LSFEM, namely, with both good stability and accuracy in solving radiative transfer in strongly inhomogeneous media. The investigation on the FEM based on the MSORTE is proceeding in a separate work.

In summary, it is demonstrated and proved that the LSFEM will suffer a deficiency problem in solving radiative transfer in strongly inhomogeneous media. This violate the traditional impression on the performance of the LSFEM. The theoretically results obtained based on the analysis of numerical methods for solving homogeneous media cannot be taken directly for strongly inhomogeneous media without thorough verification.

## 4. Numerical Tests

### *4.1. Infinite slab with an inclusion layer*

The radiative heat transfer in an inhomogeneous slab with a non-scattering inclusion layer is studied. The inclusion layer is located in the center of the slab. The thickness of the slab is $L$ and the thickness of the layer is $d = 0.4L$. The optical thickness of the inclusion layer is defined as $\tau_d = \bar{\kappa}_a d$, where $\bar{\kappa}_a$ is the absorption coefficient of the inclusion layer. The other portion of the slab is non-participating. The temperature of the left wall is $T_L$=1000K, the temperature of the inclusion layer is $T_g = 500K$, and the right wall is kept cold. Both the left and the right walls are black. In this case, the localized inclusion layer forms an inhomogeneity of medium properties, and the strength of the inhomogeneity is governed by $\tau_d$. This case was also studied in [35].

The LSFEM and the GFEM are applied to solve the heat flux distribution in the inhomogeneous slab. The dimensionless heat flux distribution $q(x)/\sigma T_L^4$ in the slab solved by the LSFEM and the GFEM are presented in Figure 8(a) and (b), respectively, for different values of $\tau_d$. During the solutions, the slab is uniformly subdivided into 100 elements and the zenith angle is discretized using the Gauss integration with 20 discrete directions. The results obtained using the discontinuous spectral element method [37] are taken as the reference results (exact).

Generally, the heat flux distributions obtained by using the LSFEM is free of spurious wiggles, and agree well with the exact results for small value of the optical thickness of the inclusion layer, say, $\tau_d$ = 0.1. However, it is seen that particularly large numerical errors are observed with the increasing of $\tau_d$ to 10. This is the deficiency problem of the LSFEM in solving radiative transfer in strongly inhomogeneous media as discussed and analyzed in Section 3.3, which is due to the LSFEM has particularly large low frequency errors in solving this kind of problems and the error increases with strength of inhomogeneity. The heat flux distributions obtained by using the GFEM shows increasingly strong spurious wiggles with the increase of $\tau_d$ (Figure 8(b)). This is due to the stability of the GFEM, or, it has very large high frequency errors as theoretically studied in Section 3.3. The increase of $\tau_d$ causes the increase of the gradient of the intensity distribution and the heat flux distribution. The large gradient will induce the solution having a wide band spectrum distribution (with large component in high frequency) in frequency domain. As such, large errors of high frequency shown as spurious wiggles will be produced by the GFEM.

A test of the grid dependence of the solved heat flux distribution solved by the LSFEM and the GFEM is shown in Figure 9 for $\tau_d = 10$. In the test, the number of elements used is increased from $N_{el}$=100 to 2000 for the LSFEM. Though the stability maintains under different grids, it is observed that it is hard to even to obtain a acceptable converged solution to the refinement of $N_{el}$=2000. At this level of refinement, even the GFEM has show superior accuracy as compared to the LSFEM. Hence the deficiency problem of the LSFEM makes it not a better choice as compared to the GFEM to solve radiative transfer in strongly inhomogeneous media.

*4.2. Radiative transfer in an inhomogeneous square medium*

In this case, a two-dimensional radiative transfer problem is considered. The configuration of the square medium is depicted in Figure 10. The definition of variables and a sample mesh (3600 triangular elements) used for FEM solution are also shown. The absorption coefficient of the medium in the enclosure is a function defined as

$$\kappa_a(x,y) = \bar{\kappa}_a U(x+y-L) \tag{54}$$

in which $\bar{\kappa}_a$ is a introduced parameter indicating the strength of absorption, $U$ is the unit step function, and $L$ is the side length of the enclosure. The definition of $\kappa_a(x,y)$ by Eq. (54) indicates that the absorption coefficient is non zero only for $x+y>L$. The medium is non-scattering and kept cold. The radiative intensity distribution for the direction $\mathbf{\Omega} = \left[\frac{1}{\sqrt{2}}, \frac{1}{\sqrt{2}}\right]$ and side length of $L=1$ is considered. The boundary condition is prescribed at the left and the bottom walls with a intensity of unity. An exact solution for this problem can be obtained as

$$I_E(x,y) = U(x+y-L)\exp\left[-\bar{\kappa}_a \frac{x+y-L}{\sqrt{2}}\right] + U(L-x-y), \qquad x,y \in [0,L]^2 \tag{55}$$

The LSFEM and the GFEM are applied to solve the radiative intensity distribution for different values of $\bar{\kappa}_a$ under the same computational condition. The solved radiative intensity distribution along the diagonal line of the square medium (depicted in Figure 10) are presented in Figure 11 and compared with the exact solution. The solved field distribution of the radiative intensity distribution for $\bar{\kappa}_a$ =10 is shown in Figure 12. During the solution, the square enclosure is subdivided into 3600 triangular elements (shown in Figure 10).

As can be seen, the spurious wiggles are observed in the solved the radiative intensity distribution by using the GFEM. For large values of $\bar{\kappa}_a$, say $\bar{\kappa}_a$ =10, the solution obtained using the GFEM seems to be totally spoiled by the spurious wiggles (Figure 12). Though the stability remains for this two-dimensional problem, the deficiency problem of the LSFEM is obvious, which cause severely accuracy degradation. By a comparison of Figure 11 with Figure 4(b) and Figure 9, it is seen that the symptom of the deficiency problem for this two-dimensional case is the same as that for the one-dimensional case, namely, the solution accuracy near the discontinuity of the extinction coefficient deteriorates significantly with the increase of $\bar{\kappa}_a$.

In the following, the spatial convergence characteristics of the LSFEM and the GFEM is studied, which will bring better understanding of the influence of the deficiency problem of the LSFEM. To facilitate the evaluation of the convergence characteristics of the methods in solving radiative transfer in media with strong inhomogeneity, a problem with homogeneous absorption coefficient is introduced. The absorption coefficient for the homogeneous problem is defined as $\kappa_a(x,y)=\bar{\kappa}_a$ and other condition is the same as the problem defined above. The exact solution for this homogeneous problem is easily obtained as

$$I_E(x,y)=U(x-y)\exp\left[-\bar{\kappa}_a\sqrt{2}y\right]+U(y-x)\exp\left[-\bar{\kappa}_a\sqrt{2}x\right],\qquad x,y\in[0,L]^2 \tag{56}$$

The spatial convergence characteristics of the LSFEM and the GFEM is studied by solving the two problems defined above for $\bar{\kappa}_a$=10. A automatic grid generator is programmed, which can generate triangular mesh of the pattern shown in Figure 10. The square enclosure is first subdivide into $M^2$ quadrilateral elements ($M$ indicates the number of subdivisions on each dimension) and then each quadrilateral elements is subdivided into 4 triangular elements, which results in $4M^2$ triangular elements. For example, by taking $M=30$, the total number of elements generated is $N_{el}=3600$. The spatial convergence characteristics is presented in Figure 13 as the relation of maximum error with the mesh refinement parameter $M$.

By a comparison of the convergence curves for the homogeneous case, the LSFEM shows obviously superior performance than the GFEM for different grid refinement. This agree with the traditional understanding. The LSFEM is stable and the result of the LSFEM is free of spurious wiggles, while the GFEM is instable and result in oscillatory solutions. Hence the LSFEM shows superior accuracy than the GFEM.

For the inhomogeneous case, it is seen that the accuracy of both the methods deteriorate significantly and the LSFEM shows nearly the same (even worse) performance than the GFEM with grid refinement. The maximum error of result obtained by the LSFEM is still about 0.3 (the relative error is 30%) even for a fine refinement of $M=30$ $(N_{el}=3600)$. For this case, the LSFEM is still stable and the result of the GFEM is still full of spurious wiggles. The severe accuracy degradation of the GFEM can be understood as the causing of the strong spurious wiggles (such as that shown in Figure 11 and Figure 13). However, the severe accuracy degradation of the LSFEM differs the traditional understanding of the method. This severe accuracy degradation is understood to be caused by the deficiency problem of the LSFEM in solving radiative transfer in strongly inhomogeneous media, which has been theoretically analyzed in Section 3.3.

The deficiency problem of the LSFEM makes it not a reasonably better choice than the GFEM to solve radiative transfer in strongly inhomogeneous media though the latter may be instable.

## 5. Conclusions

The present work demonstrates numerically and proved theoretically that the LSFEM suffers a deficiency problem for solving radiative transfer in media with strong inhomogeneity. This deficiency problem of the LSFEM will cause a severe accuracy degradation, which compromises too much of the performance of the LSFEM. This violates the traditional impression on the performance of the LSFEM. It is also proved theoretically that the LSFEM is equivalent to a second order form of radiative transfer equation discretized by the central difference scheme.

The accuracy and stability of the LSFEM and the GFEM for solving radiative transfer in homogeneous and inhomogeneous media are studied theoretically via a developed frequency domain technique. The theoretical relations of solution error in frequency domain indicate the following facts:

(1) The GFEM has large high frequency errors for solving radiative transfer in strongly inhomogeneous media as well as for solving radiative transfer in homogeneous media. This interprets the spurious wiggles in the numerical results of the GFEM.

(2) The LSFEM has much lower high frequency errors as compared to the GFEM for solving radiative transfer both in homogeneous and in strongly inhomogeneous media. This interprets the good stability of LSFEM, namely, the results are free of ‘wiggles’.

(3) The LSFEM shows very large low frequency errors for solving radiative transfer in strongly inhomogeneous media. This interprets the deficiency problem of the LSFEM, namely, the severe accuracy degradation for solving radiative transfer in strongly inhomogeneous media.

(4) The error distribution of the LSFEM is very similar to that of the LSORTE discretized by central difference scheme. This confirms the equivalence of the LSFEM discretization with the LSORTE [35] discretized by central difference scheme.

Because of the deficiency problem, the LSFEM is not reasonably a better choice than the GFEM to solve radiative transfer in strongly inhomogeneous media. As such, a scheme of FEM that is both accurate and stable for solving radiative transfer in strongly inhomogeneous media is still appealing.

## Acknowledgements

This work is supported by the National Nature Science Foundation of China (Nos. 50906017, 50836002). The first author also thanks the support of the Development Program for Outstanding Young Teachers in Harbin Institute of Technology (HITQNJS.2009.020).

## Figure Captions

**Figure 1.** Schematic of the 1D grid system and the definition of variables.

**Figure 2.** The frequency domain distribution of solution error for the LSFEM with comparing with the GFEM at different grid optical thickness.

**Figure 3.** Relation of the solution error with grid optical thickness for the LSFEM and the GFEM reduced frequency $\bar{\varpi} = 0.5$.

**Figure 4.** Solved intensity distribution in the inhomogeneous slab by using the GFEM and the LSFEM for different extinction function distributions, **(a)** GFEM and **(b)** LSFEM.

**Figure 5.** Solved intensity distribution in the inhomogeneous slab by the GFEM with and without using the source-linearization technique for different extinction functions, **(a)** $\bar{\kappa}_a = 0.1$, **(b)** $\bar{\kappa}_a = 1$ and **(c)** $\bar{\kappa}_a = 2$.

**Figure 6.** Solved intensity distribution at $\bar{\kappa}_a = 1$ by using the meshless method and the FEM with source-linearization formulation: **(a)** meshless method based on the RTE, **(b)** meshless method based on the LSORTE and the MSORTE, **(c)** FEM based on the RTE, LSORTE and the MSORTE.

**Figure 7.** Frequency domain error distribution of the RTE, LSORTE and MSORTE discretized by central difference scheme based on source-linearization formulation.

**Figure 8.** Solved dimensionless heat flux distribution in the inhomogeneous slab by different methods, (a) LSFEM, (b) GFEM.

**Figure 9.** Grid dependence of the solved heat flux distribution solved by the LSFEM and the GFEM.

**Figure 10.** Schematic of the square solution domain and definition of variables (a sample of mesh used for FEM solution is also shown, 3600 triangular elements).

**Figure 11.** Intensity distribution along the diagonal line of the square enclosure ($y = x$) solved by the LSFEM and the GFEM.

**Figure 12.** Radiative intensity field solved by the GFEM and the LSFEM.

**Figure 13.** Spatial convergence characteristics of the LSFEM and the GFEM for solving the case of $\bar{\kappa}_a = 10$ (the number of elements is $N_{el} = 4M^2$).

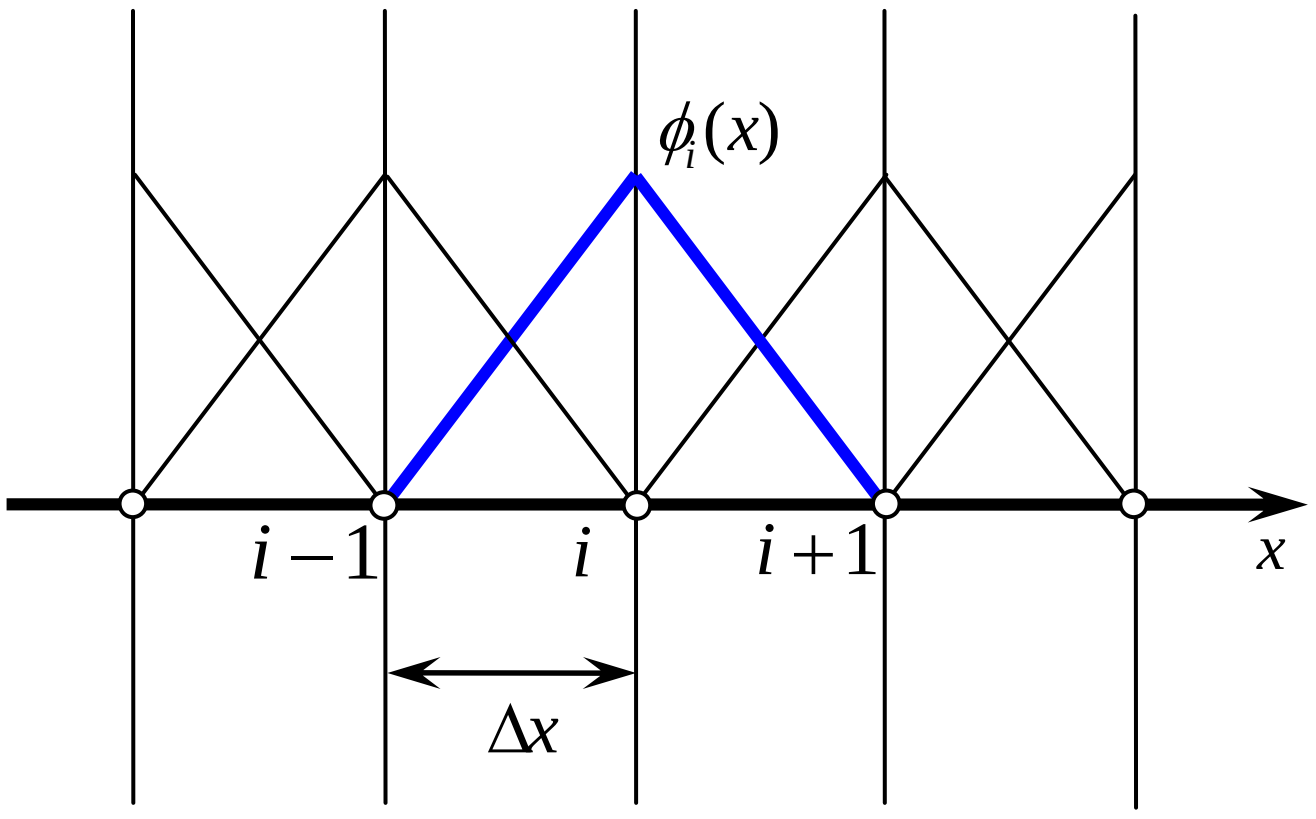

**Figure 1**

**Authors: Zhao, Tan and Liu**

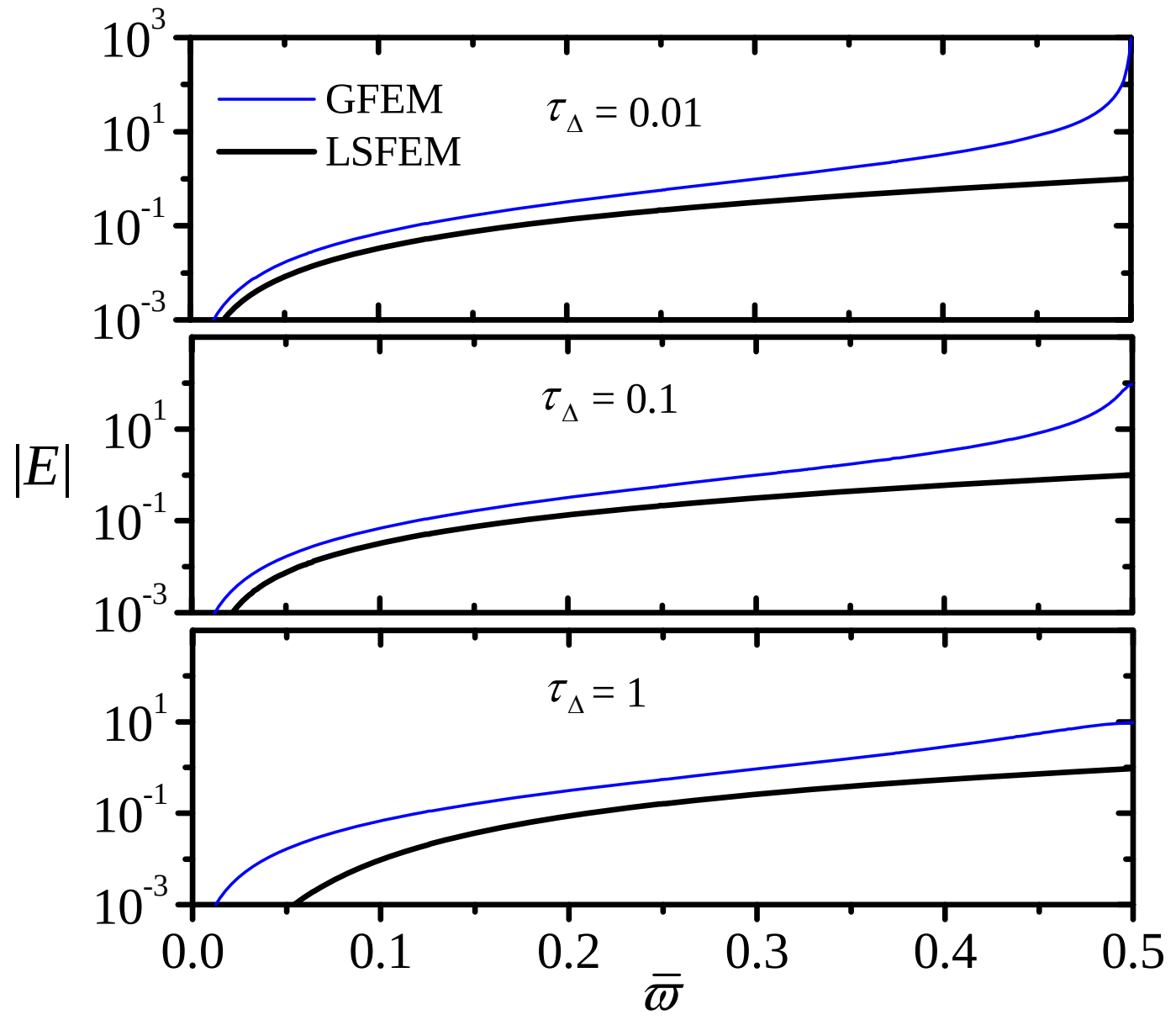

**Figure 2**

**Authors: Zhao, Tan and Liu**

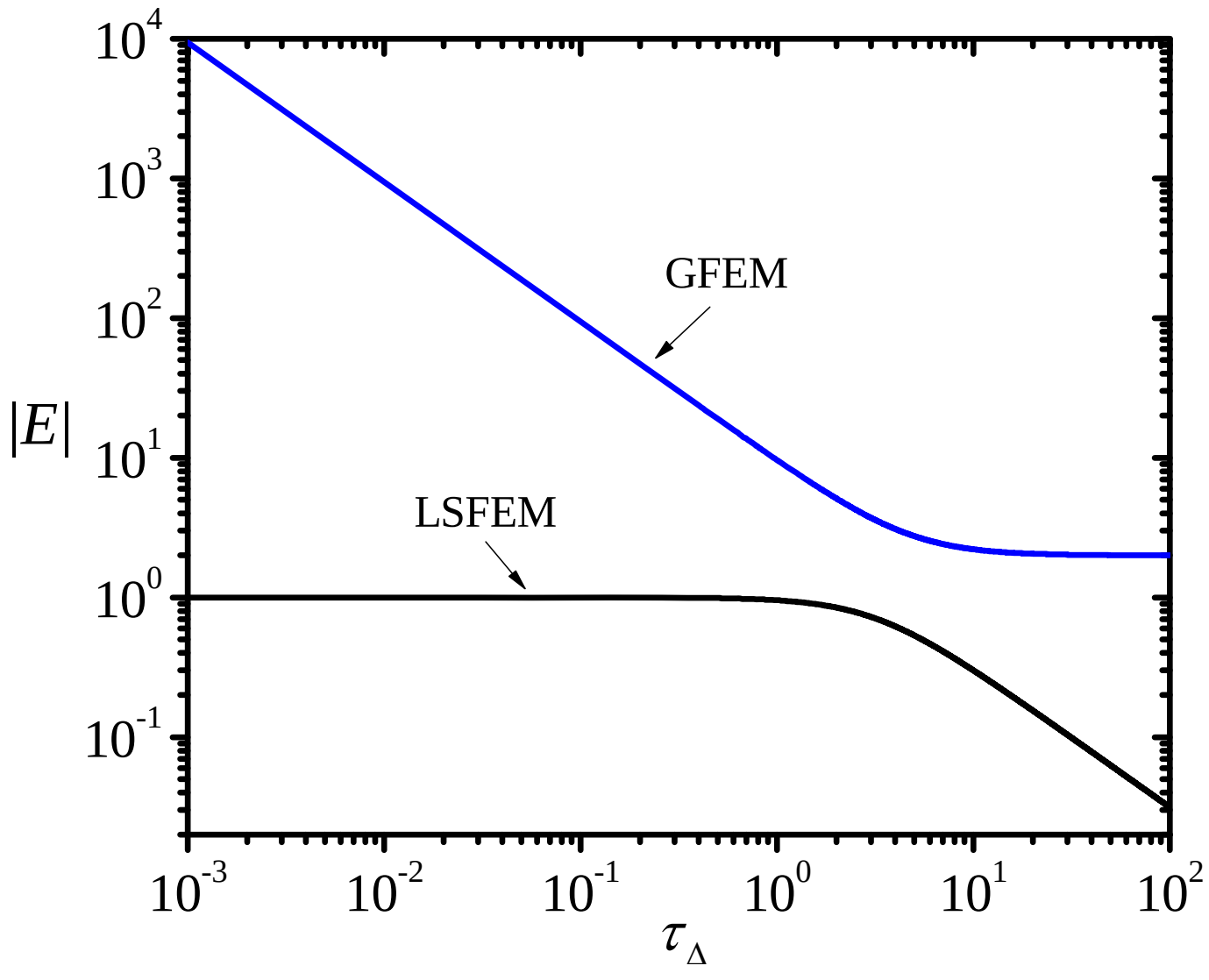

**Figure 3**

**Authors: Zhao, Tan and Liu**

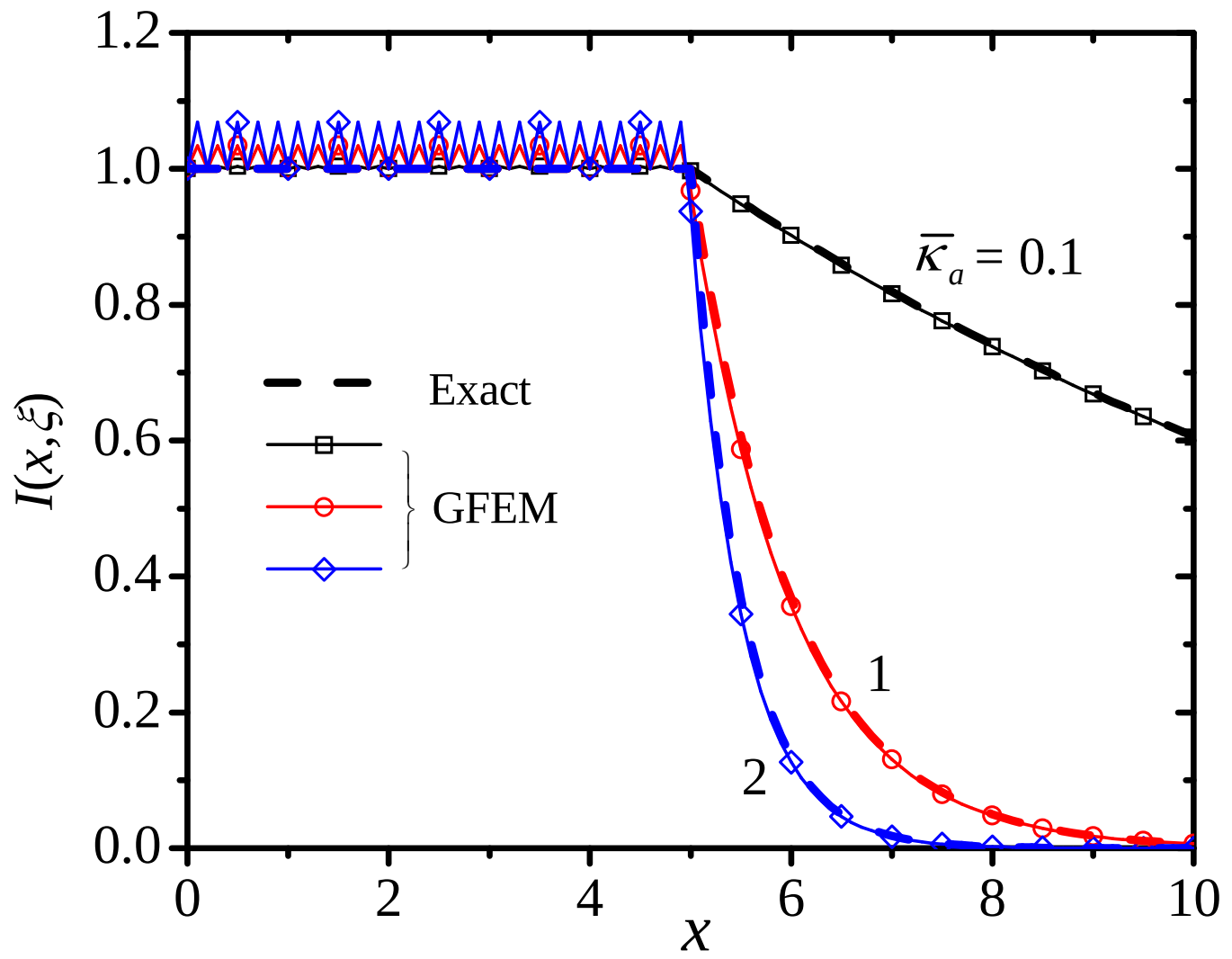

**Figure 4(a)**

**Authors: Zhao, Tan and Liu**

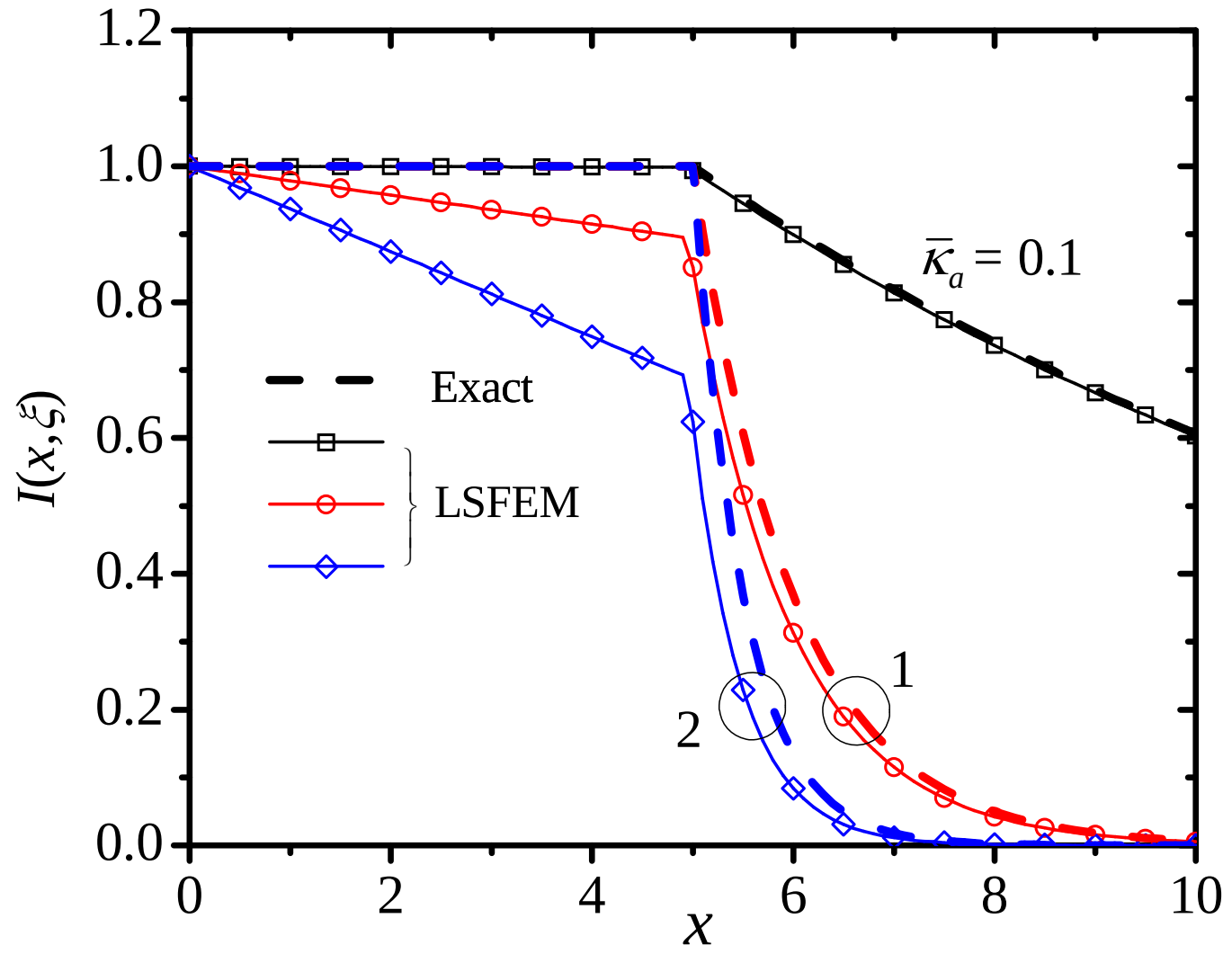

**Figure 4(b)**

**Authors: Zhao, Tan and Liu**

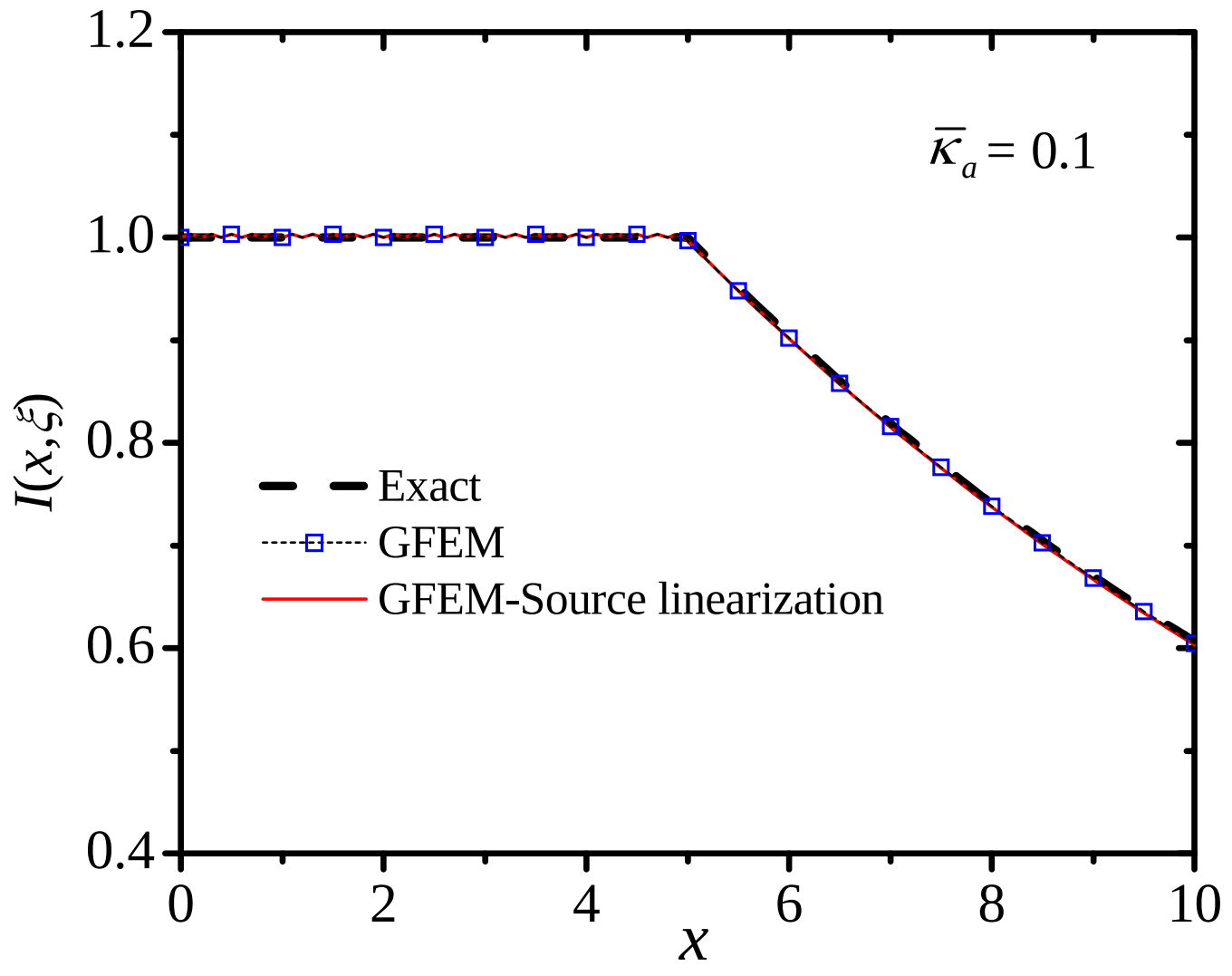

**Figure 5(a)**

**Authors: Zhao, Tan and Liu**

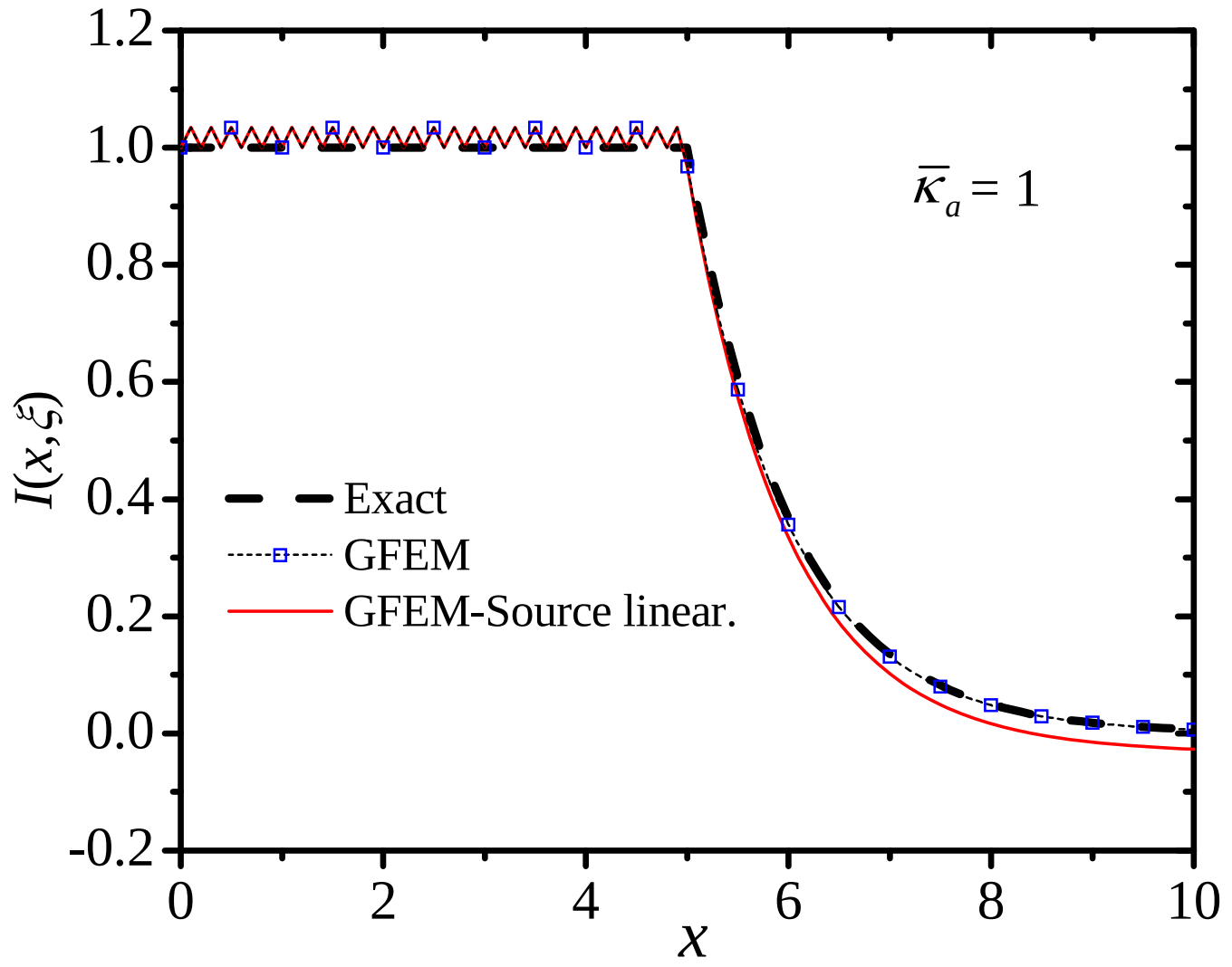

**Figure 5(b)**

**Authors: Zhao, Tan and Liu**

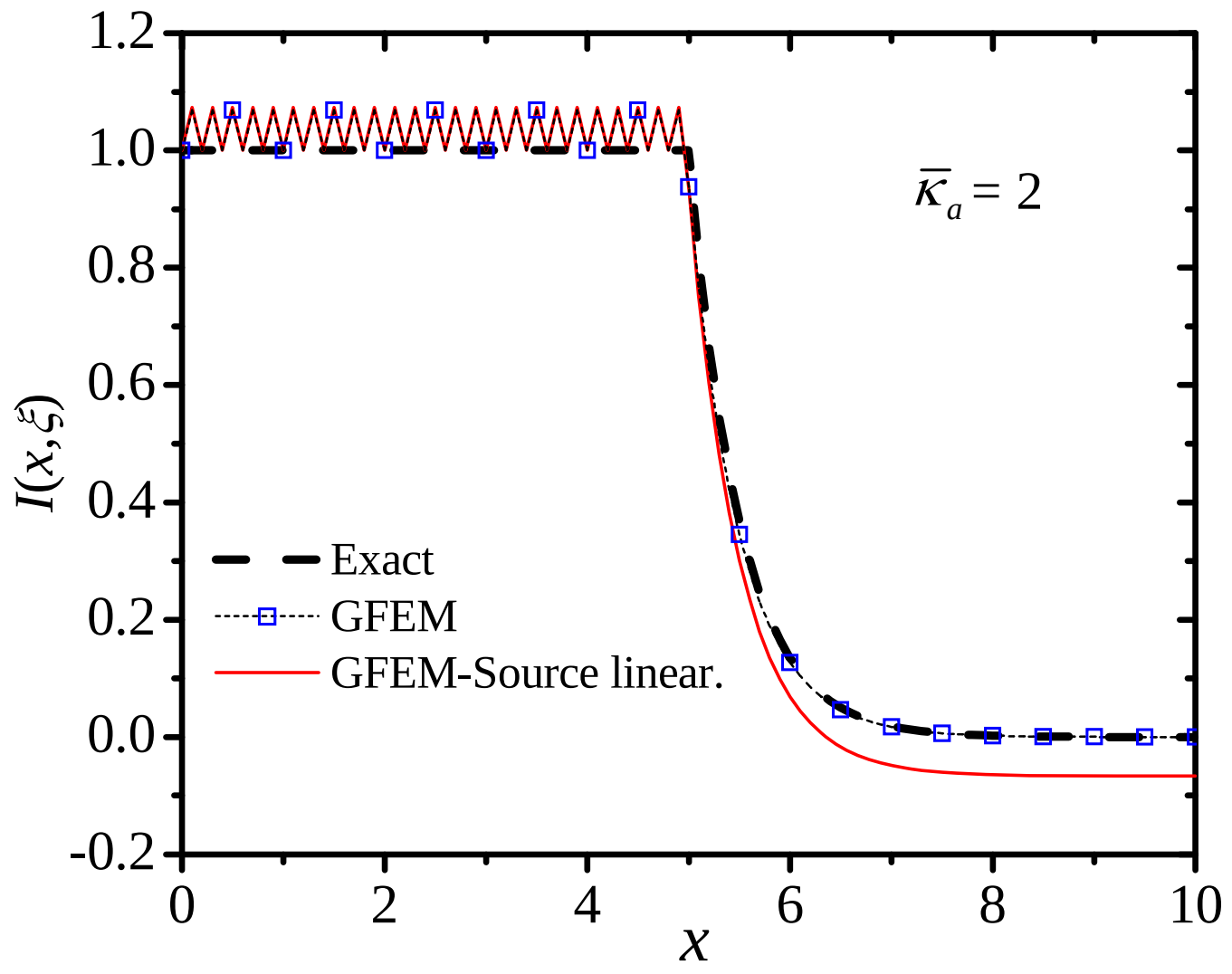

**Figure 5(c)**

**Authors: Zhao, Tan and Liu**

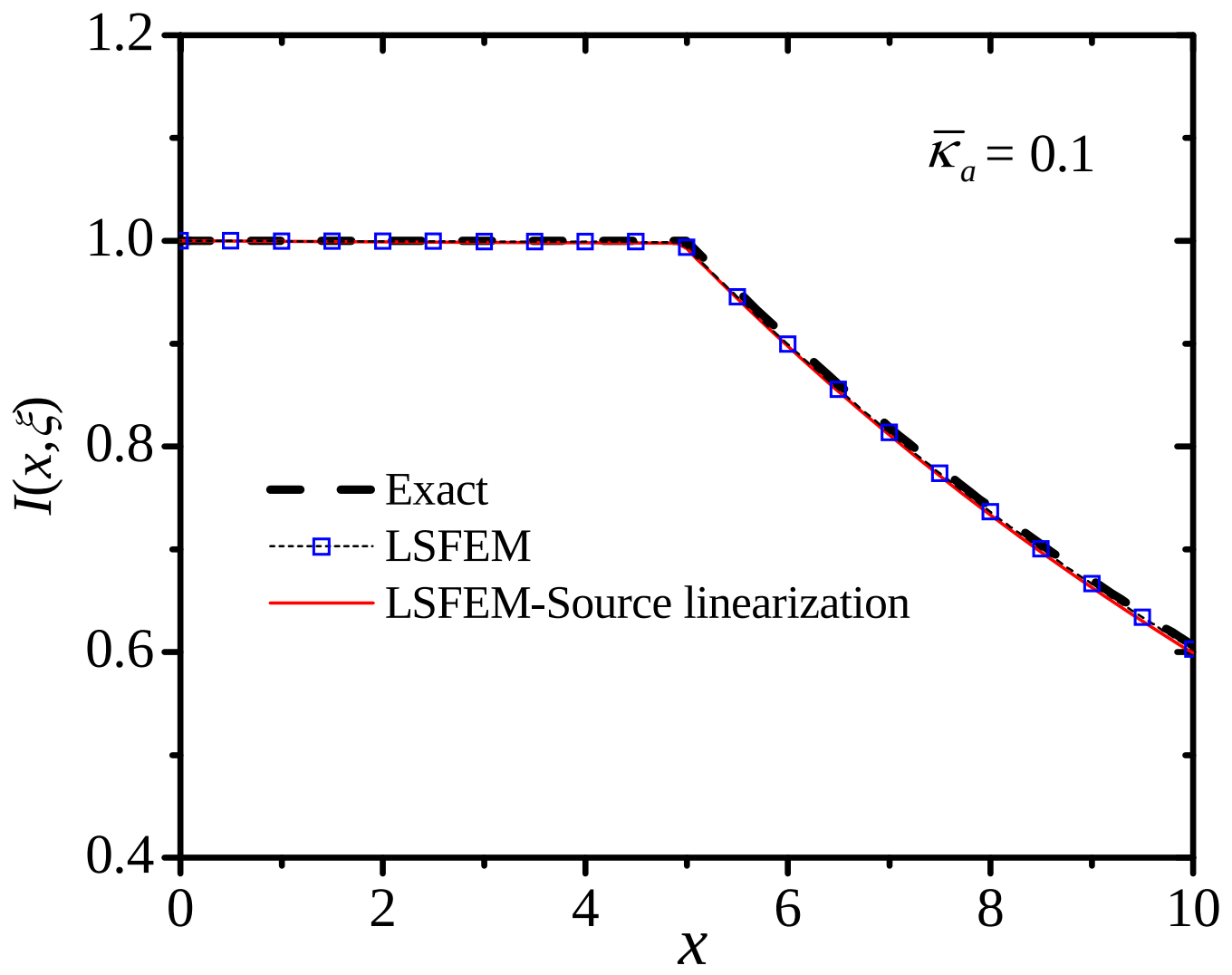

**Figure 6(a)**

**Authors: Zhao, Tan and Liu**

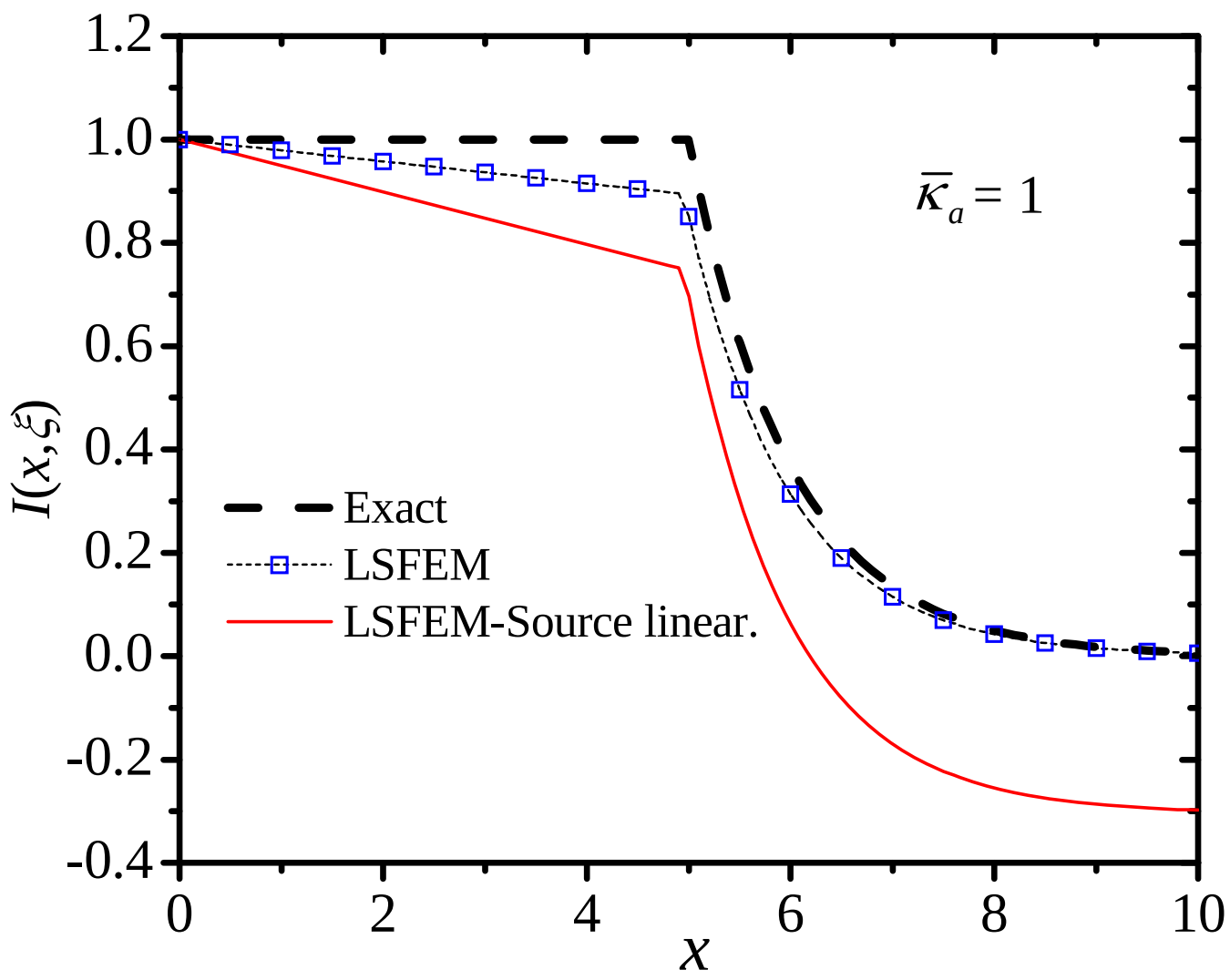

**Figure 6(b)**

**Authors: Zhao, Tan and Liu**

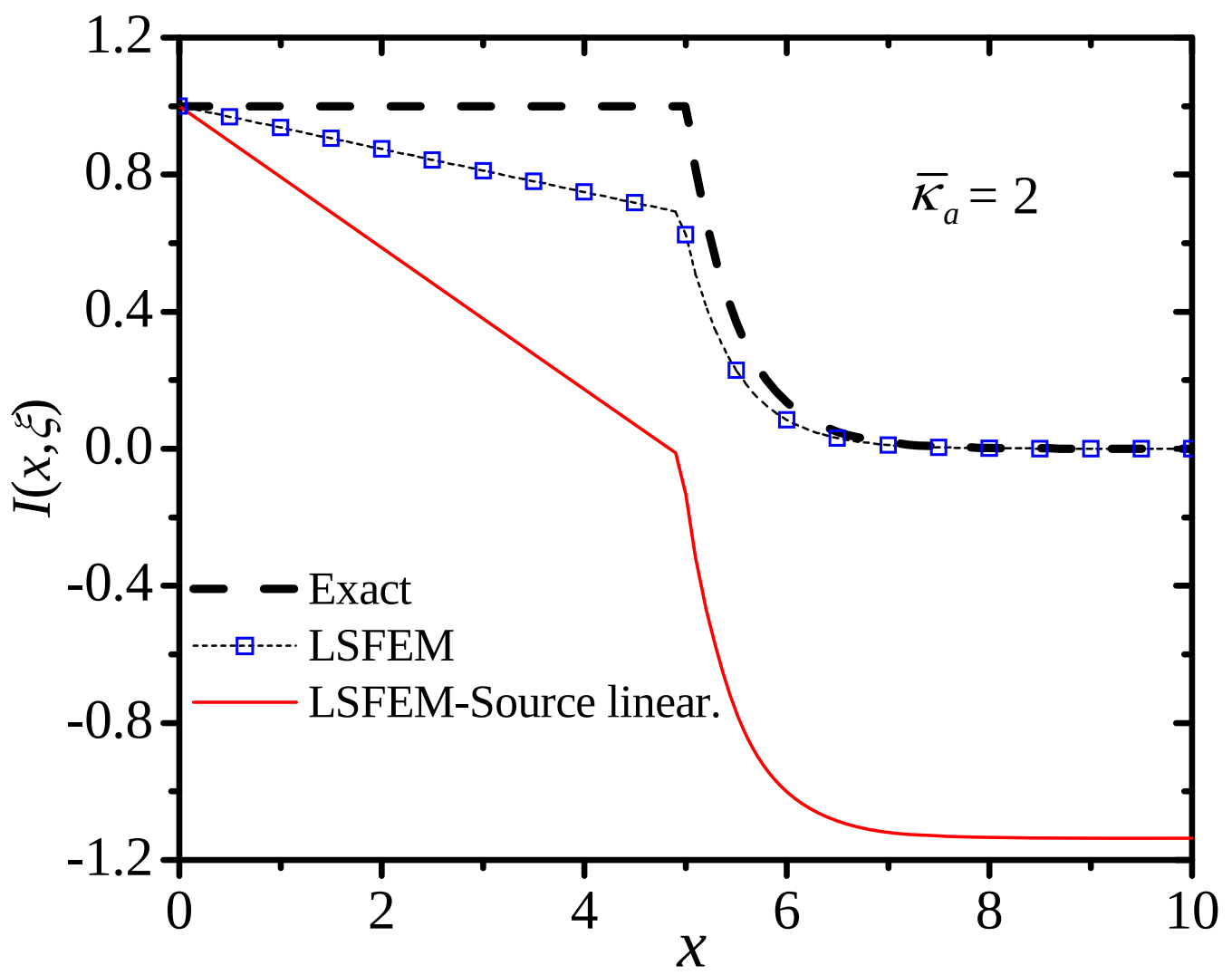

**Figure 6(c)**

**Authors: Zhao, Tan and Liu**

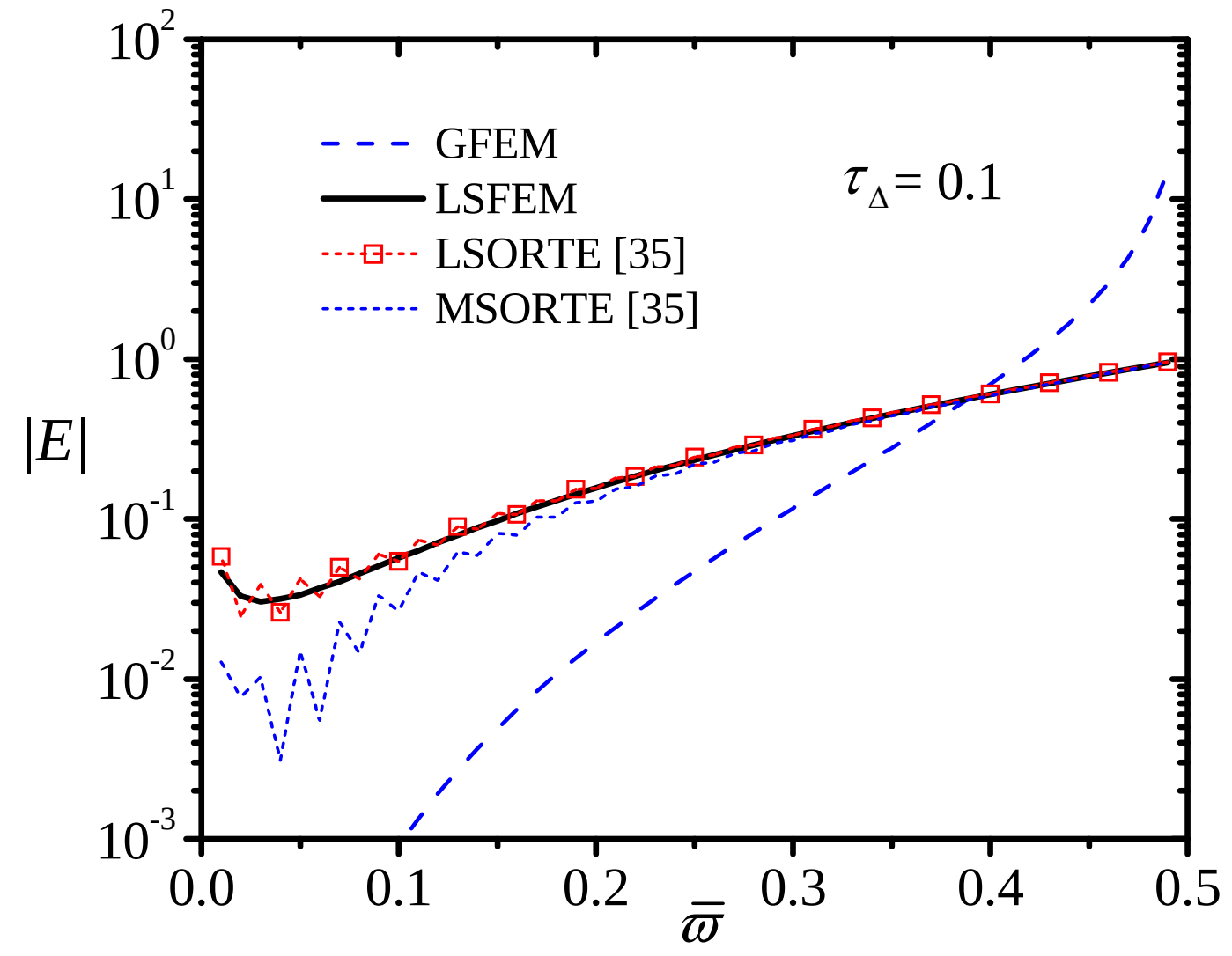

**Figure 7(a)**

**Authors: Zhao, Tan and Liu**

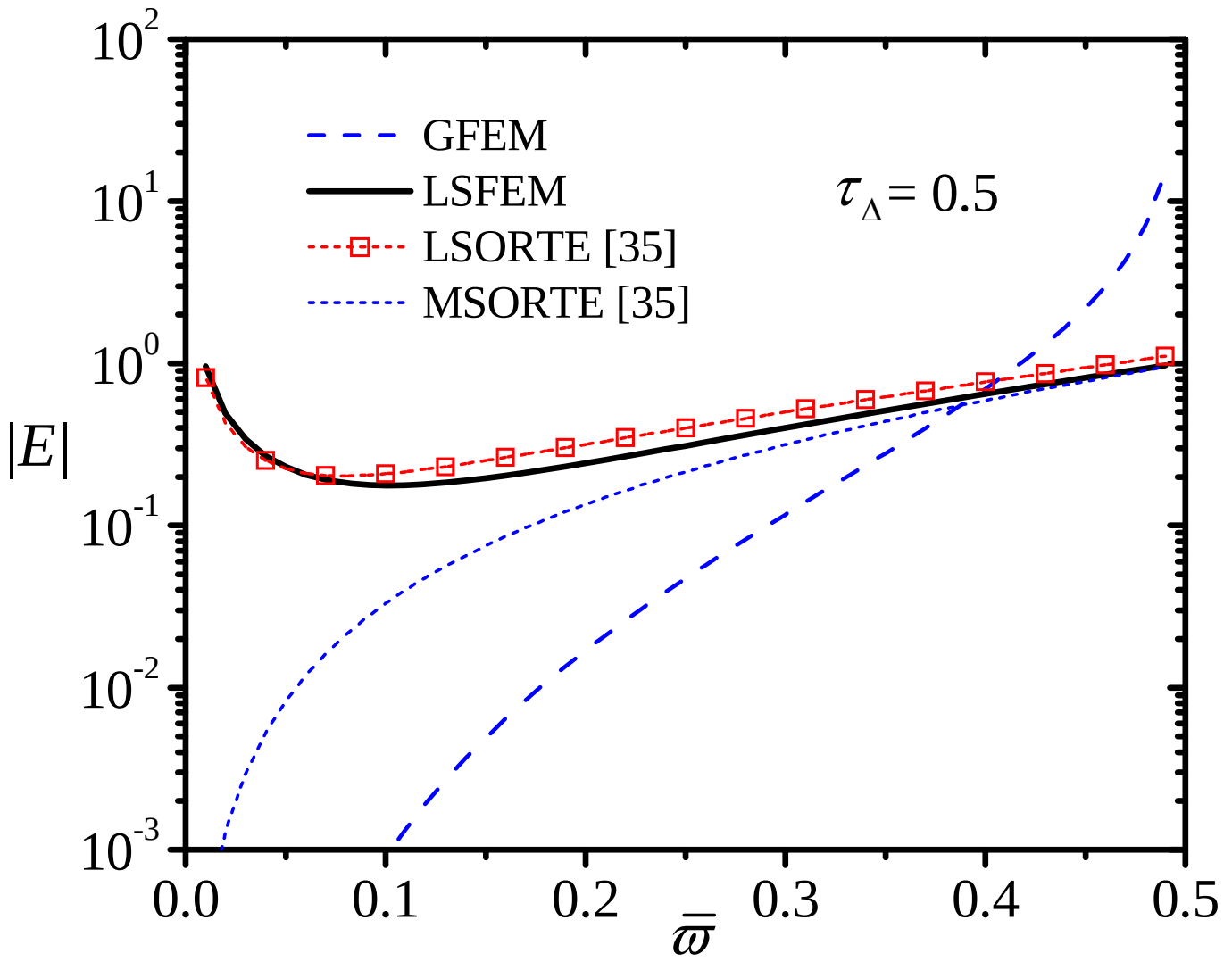

**Figure 7(b)**

**Authors: Zhao, Tan and Liu**

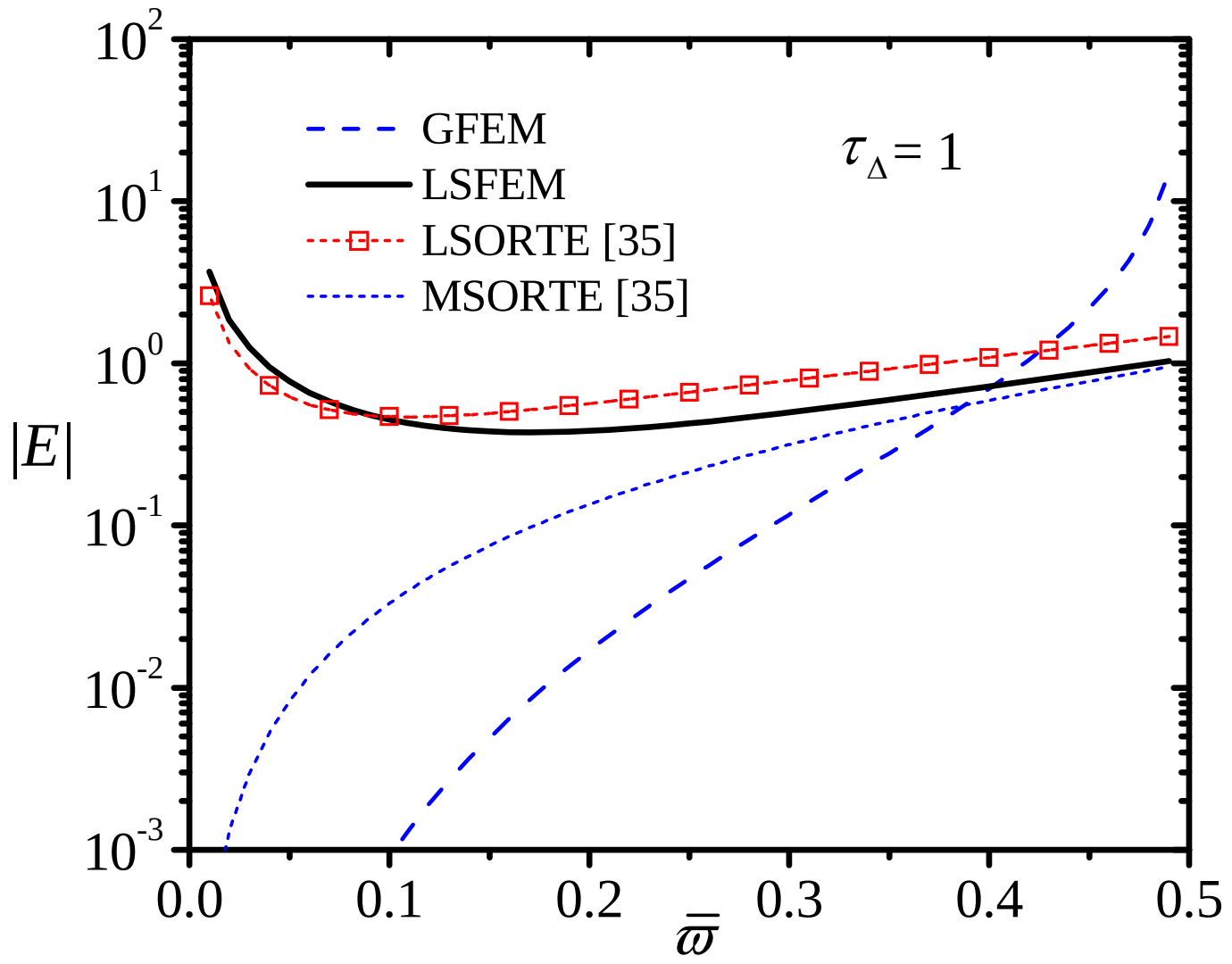

**Figure 7(c)**

**Authors: Zhao, Tan and Liu**

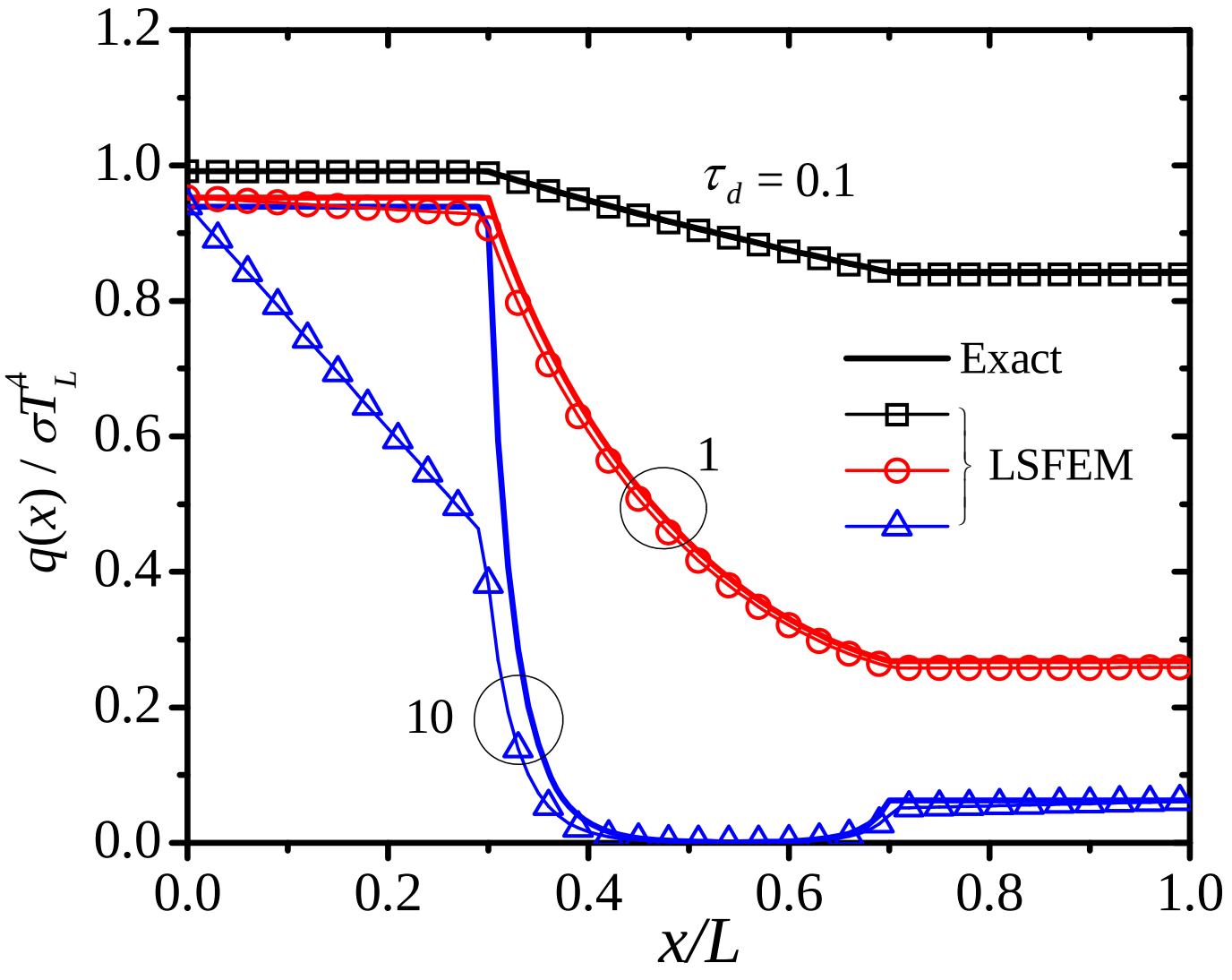

**Figure 8(a)**

**Authors: Zhao, Tan and Liu**

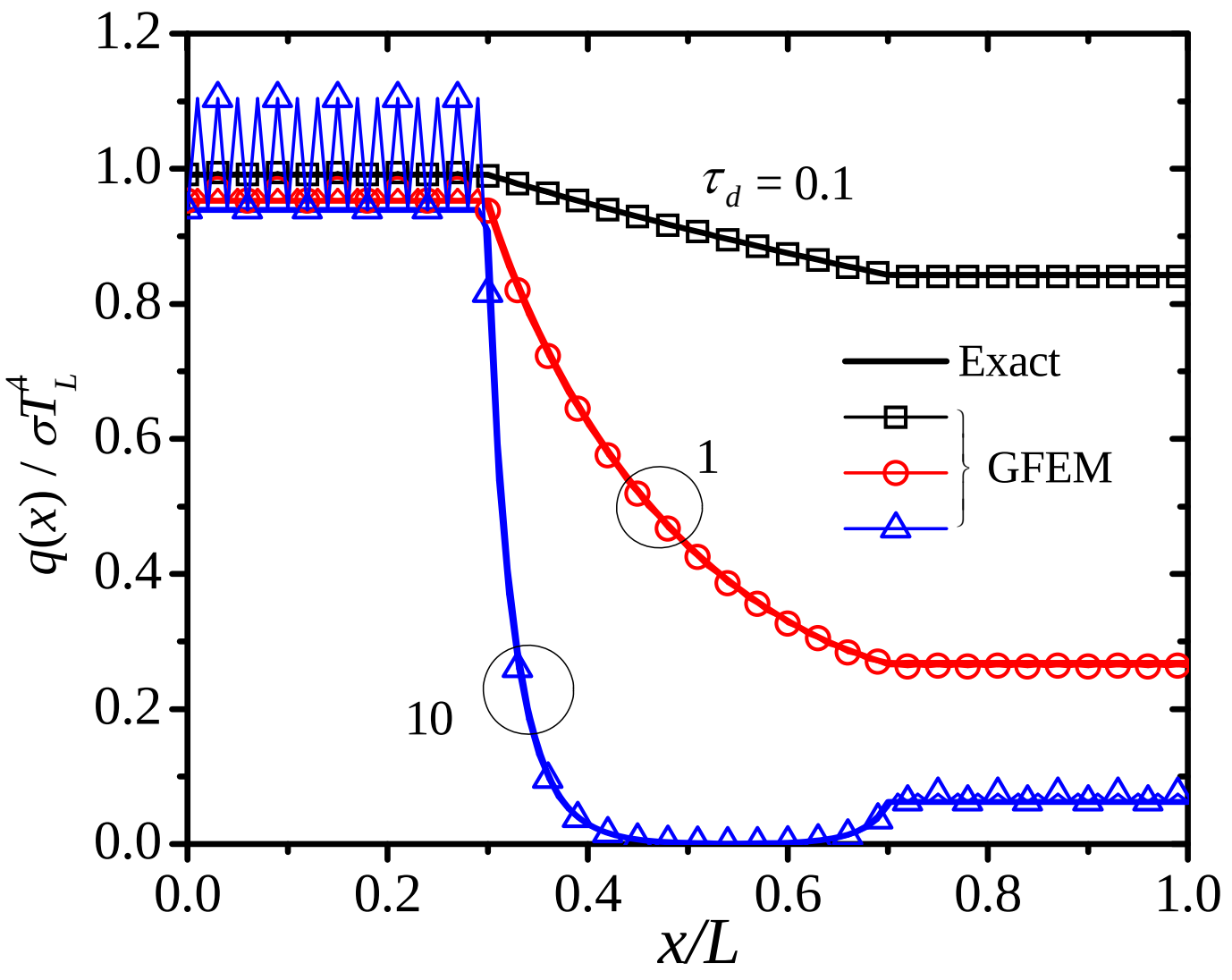

**Figure 8(b)**

**Authors: Zhao, Tan and Liu**

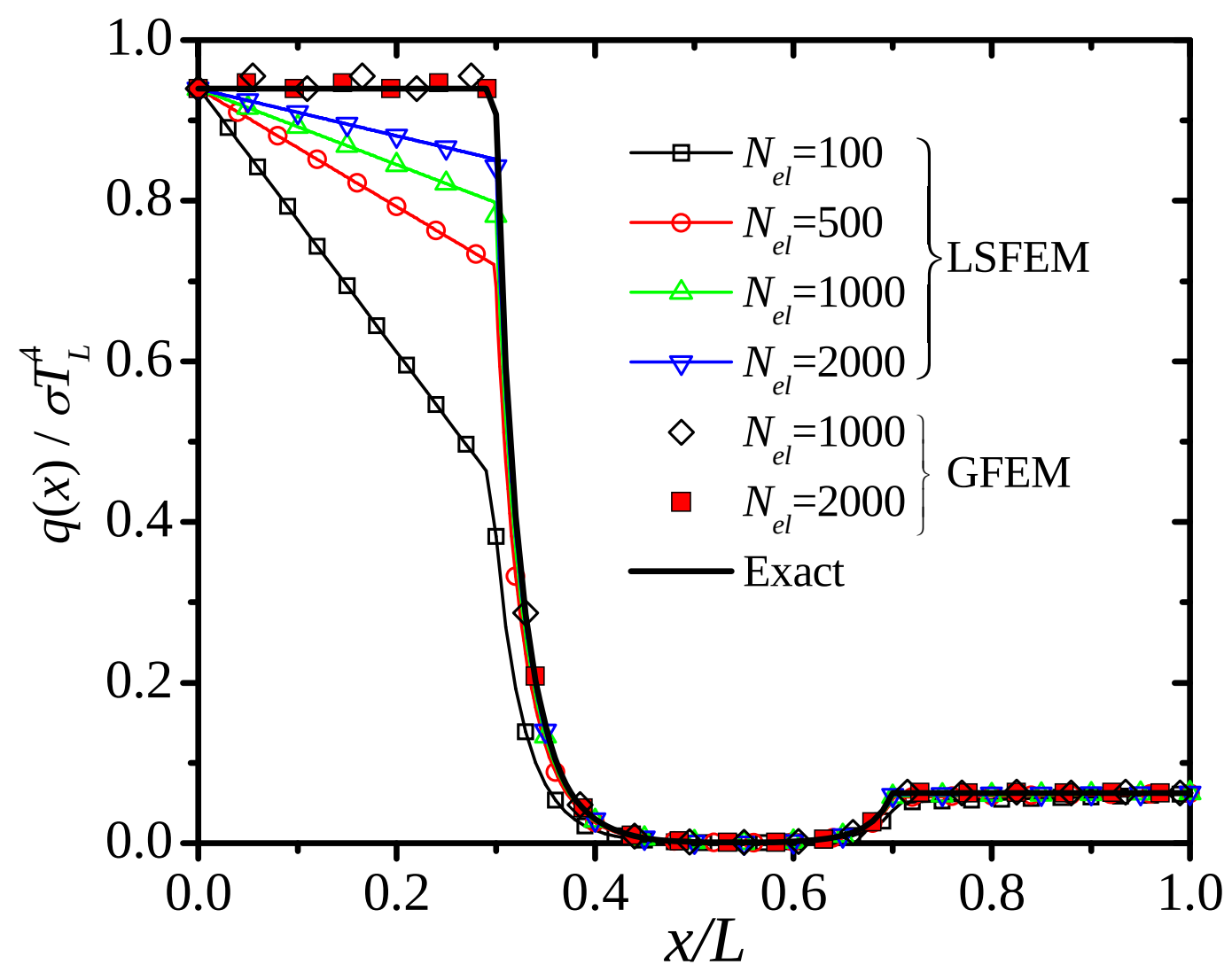

**Figure 9**

**Authors: Zhao, Tan and Liu**

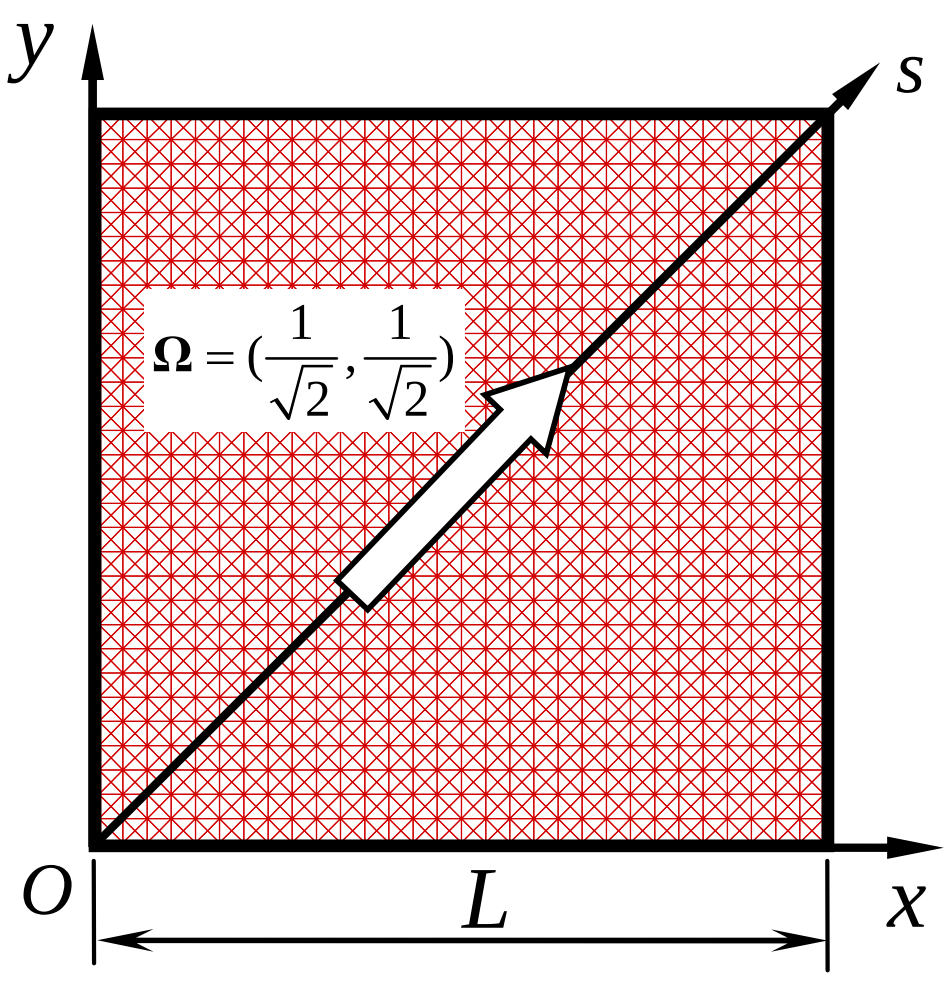

**Figure 10**

**Authors: Zhao, Tan and Liu**

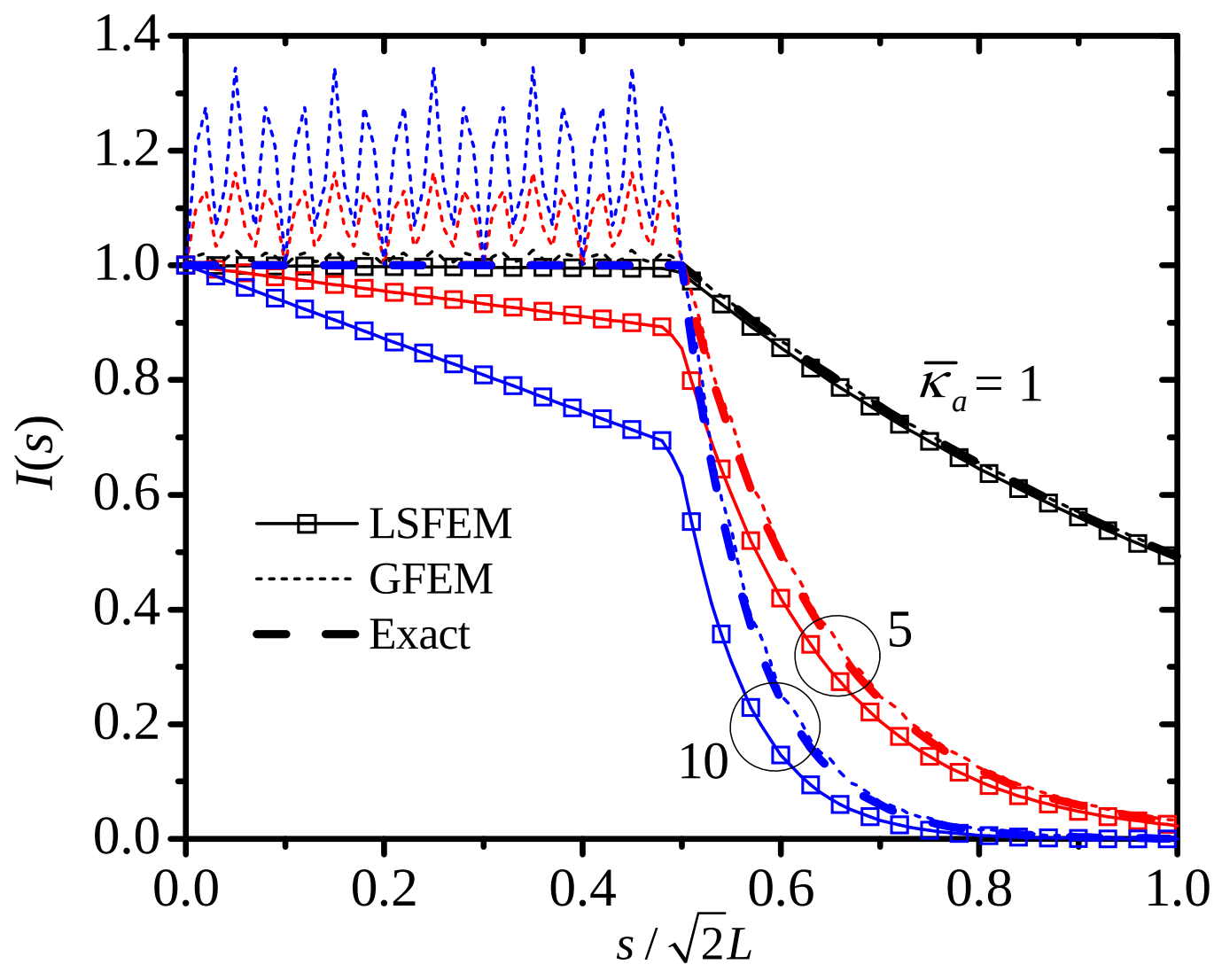

**Figure 11**

**Authors: Zhao, Tan and Liu**

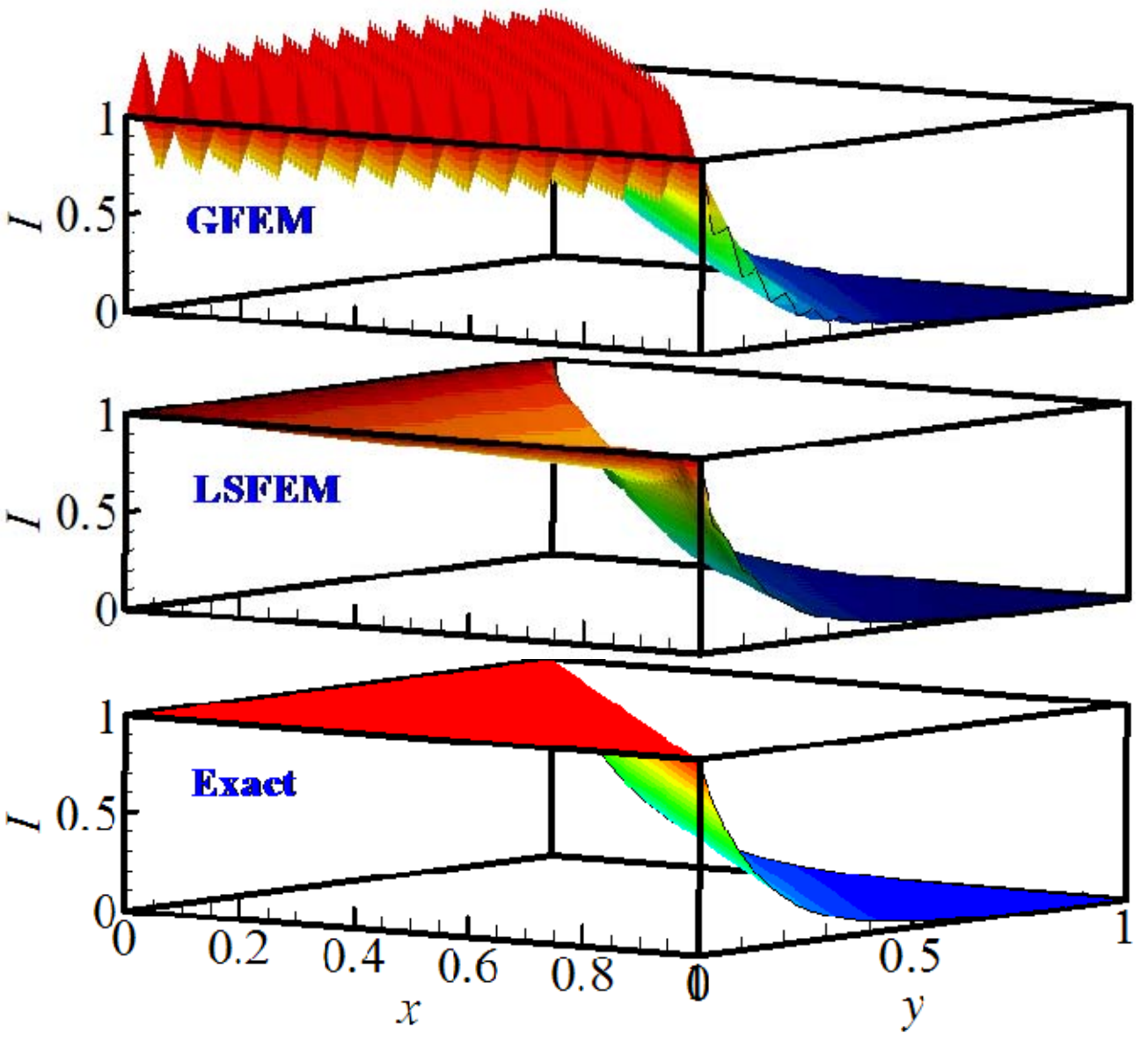

**Figure 12**

**Authors: Zhao, Tan and Liu**

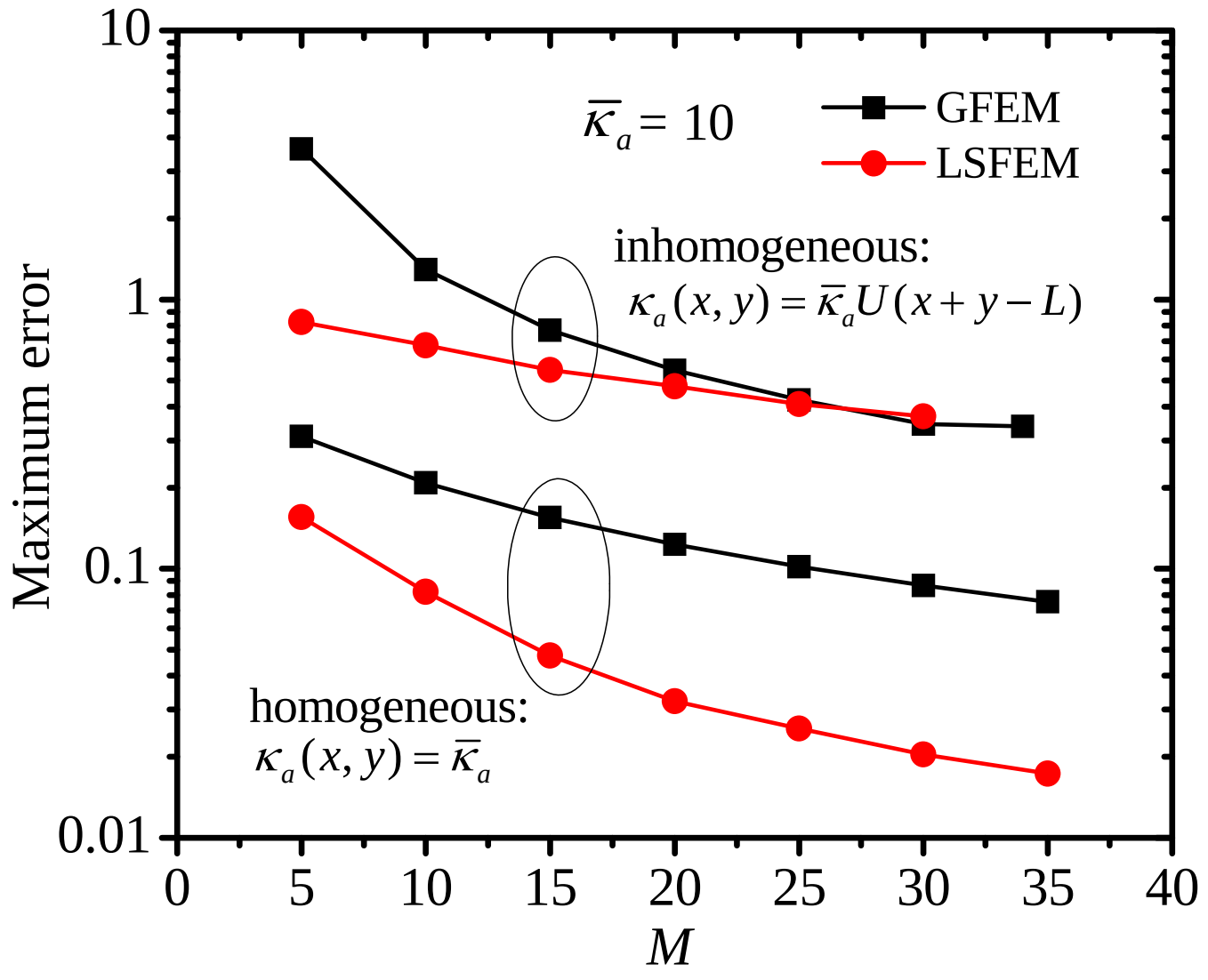

**Figure 13**

**Authors: Zhao, Tan and Liu**